\documentclass[fleqn,usenatbib]{mnras}

\usepackage[T1]{fontenc}
\usepackage{ae,aecompl}

\usepackage{graphicx}	
\usepackage{newtxtext,newtxmath}
\usepackage{comment}
\usepackage{orcidlink}

\newcommand{\rc}{\color{black}}
\newcommand{\rcc}{\color{black}}

\title[Aerosol dynamics on hot exoplanets]{Aerosol dynamics on hot exoplanets: the role of radiation pressure}

\author[Owen, J. E. \& Murray-Clay, R. A.]{
James E. Owen$^{\orcidlink{0000-0002-4856-7837}1,2}$\thanks{E-mail: james.owen@imperial.ac.uk}
and
Ruth A. Murray-Clay$^{\orcidlink{0000-0001-5061-0462}3}$\\
$^{1}$Imperial Astrophysics, Imperial College London, Blackett Laboratory, Prince Consort Road, London SW7 2AZ, UK\\
$^{2}$Department of Earth, Planetary, and Space Sciences, University of California, Los Angeles, CA 90095, USA\\
$^{3}$Department of Astronomy and Astrophysics, University of California, Santa Cruz, CA 95064, USA
}

\date{Accepted XXX. Received YYY; in original form ZZZ}

\pubyear{2015}

\begin{document}
\label{firstpage}
\pagerange{\pageref{firstpage}--\pageref{lastpage}}
\maketitle

\begin{abstract}
Aerosols appear to be ubiquitous in exoplanetary atmospheres. However because our understanding of the physical processes that govern aerosols is incomplete, their presence makes the measurement of atmospheric properties, such as molecular abundance ratios, difficult. We show that aerosol particles in highly-irradiated exoplanets experience an additional acceleration due to stellar radiation pressure. The strength of this radiative acceleration often exceeds the planet's gravity and can approach values of $\sim10-20\times$ gravity's for low-density planets (typically sub-Saturns) hosting $\sim$0.1--1$\mu$m aerosols. Since these highly irradiated, low-density planets are often the best targets for atmospheric characterisation with current instrumentation, radiation pressure is likely an important process when considering aerosol dynamics. We find that radiation pressure accelerates hazes produced by photochemistry at high altitudes to faster terminal velocities, causing them to grow more slowly. Hence, the particles are smaller and have lower {\rc mass} concentrations in the presence of radiation pressure. By simulating haze-like aerosols in a 2D equatorial band model, we show that radiation pressure steepens optical slopes in transmission spectra, {\rc resulting in} less muted molecular features in the Near-IR and gives rise to a correlation between the strength of radiation pressure and the molecular feature amplitude.  {\rcc Furthermore, the interaction of zonal winds and radiation pressure impacts both the optical slopes and amplitudes on the individual morning and evening terminators.}
\end{abstract}




\section{Introduction}
Aerosols are ubiquitous in planetary atmospheres, from Solar System bodies to exoplanets \citep[e.g.][]{Horst2018,Helling2019,Gao2021}.  Their presence in highly-irradiated planets was postulated shortly after the discovery of the first hot jupiters \citep[e.g.][]{Guillot1996}, and it was further demonstrated that aerosols could shape and modify the observables of an exoplanet's atmosphere \citep[e.g.][]{Burrows1997,Seager2000a,Seager2000b,Barman2001}. 

 The slant optical path that arises in transmission spectroscopy means that this method is particularly sensitive to aerosols \citep[e.g.][]{Fortney2005}. In fact, the first transmission spectrum of an exoplanet showed strong evidence of aerosols at sufficiently high altitudes to impact the 5893~\AA~ Sodium doublet \citep{Charbonneau2002}. Optical transmission spectra, mainly of hot Jupiters, have revealed the impact of small $\sim 0.1~\mu$m size particles at millibar pressures \citep[e.g.][]{Lecavelier2008,Wakeford2017,Benneke2019}, with a wide variety of slopes and feature amplitudes \citep[e.g.][]{Sing2016}. To complicate matters, the origin of aerosols remains uncertain, with condensate particles --``clouds''-- and photochemically produced particles --``hazes''-- both thought to exist in exoplanet atmospheres \citep[e.g.][]{Gao2021}\footnote{We are following the nomenclature suggested by \citet{Gao2021}, based on the discussion of \citet{horst_clouds_2016}.}.

Near-IR transmission spectroscopy, mainly with HST, provided the first sample of the properties of exoplanet atmospheres. Early observations of the 1.4$\mu$m water feature showed that it was smaller than predicted by aerosol-free models \citep[e.g.][]{Deming2013,Sing2013}, indicating the presence of aerosols. When observed in transmission, the impact on molecular features is that aerosols typically obscure all stellar light below a certain pressure level across a broad wavelength range. Thus, instead of observing the full amplitude of a molecular feature, it appears ``muted'' with only the regions of highest opacity in the feature appearing in the spectrum \citep[e.g.][]{Fortney2005}. Statistical studies of the amplitude of the 1.4$\mu$m water feature have revealed a wide variety of spectral amplitudes \citep[e.g.][]{Fu2017,Dymont2022} with planets such as GJ1214 b providing no evidence of an absorption feature \citep[e.g.][]{Kreidberg2014}. In contrast, other planets show large-amplitude absorption features \citep[e.g.][]{Fraine2014,Nikolov2016,Tsiaras2018}. These results have continued with the advent of JWST \citep[e.g.][]{Feinstein2023,Madhusudhan2023,Kempton2023}, which has also provided the first direct evidence for an absorption feature from cloud particles \citep{Grant2023}.

Retrieval studies, which attempt to incorporate clouds, have tried to investigate trends with planetary parameters \citep[e.g.][]{Barstow2017,Pinhas2019}; however, these analyses are impacted by the details of the aerosol parameterisations used in the retrieval models \citep[e.g.][]{Barstow2020}, and limited wavelength coverage \citep[e.g.][]{Lueber2024}. Thus, we need a complete picture of the physics controlling aerosols in exoplanet atmospheres to progress. One complication is that while transmission spectra often demonstrate the ubiquity of aerosols, albedo measurements and emission spectra of many hot Jupiters show evidence for gas absorption and emission without a large amount of aerosols \citep[e.g.][]{Heng2013,Line2014, Barstow2014}, with a few notable exceptions where the albedo is large \citep[e.g.][]{Heng2021,Adams2022}. However, there is no clear correlation with planetary or stellar parameters yet. The explanation for these longitudinal variations is not entirely clear, with many models unable to characterise the combined observation sets \citep[e.g.][]{Gao2021}.

First principled microphysical models have been developed to study the formation, evolution, dynamics and observational impact of aerosols on exoplanet atmospheres \citep[e.g.][]{Helling2008,Gao2018,Kawashima2019}, as well as more simplified schemes that parameterise different aspects of the physics, including nucleation, particle size distribution, and radiative transfer \citep[e.g.][]{Ohno2017,Ormel2019}. Furthermore, aerosol particles have been inserted into 3D GCMs \citep[e.g.][]{Parmentier2013,Komacek2019} or post-processed with more sophisticated 1D models \citep[e.g.][]{Lee2015,Parmentier2016,Tan2017}. These models have been applied to observations with a variety of successes \citep[e.g.][]{Parmentier2016,Roman2021}; however, the degeneracies in the observations and incomplete physical understanding mean that the parameters that drive the presence, properties, and observable impact of aerosols are still poorly understood \citep[e.g.][]{Dymont2022}.  

Of particular note are the often steep optical slopes observed in transmission of some planets \citep[e.g.][]{Pinhas2019,Welbanks2019}, steeper than can be explained by standard models. Stellar activity could be the explanation in some of these cases \citep[e.g.][]{Espinoza2019}, and \citet{Pinhas2017} showed that sulphide condensates could be the cause in other scenarios. Alternatively, \citet{Ohno2020} proposed that extremely efficient vertical transport could produce these steep slopes; however, the strength of the vertical transport required can exceed that predicted by GCMs \citep{Zhang2018}. 

Therefore, it is clear that our understanding of the physical processes underlying the dynamics of exoplanetary aerosols remains incomplete, with several observational characteristics still to be understood. In this work, we demonstrate an additional physical process that is yet to be studied in hot exoplanets that is likely to be essential for controlling aerosol dynamics and, hence, their observable properties: stellar radiation pressure.

Stellar radiation pressure has long been known to be influential on small particles in astrophysics \citep[e.g.][]{Burns1979}, where radiation pressure can unbind them from their host star. In the presence of gas drag, the interactions become richer \citep[e.g.][]{Takeuchi2003,OK2019} resulting in an interplay between gravity, drag, and radiation pressure. Here, we show that the strength of radiation pressure on aerosols at altitudes relevant for atmospheric observations can often exceed that of the planet's gravity for hot exoplanets. First, we discuss the general impact of radiation pressure on aerosol particles. Then, we focus on hazes, where we analytically and numerically study radiation pressure's impact on the dynamics and observational properties of aerosol particles.

\section{Overview: the importance of radiation pressure}\label{sec:overview}
\begin{figure}
    \centering
    \includegraphics[width=\columnwidth]{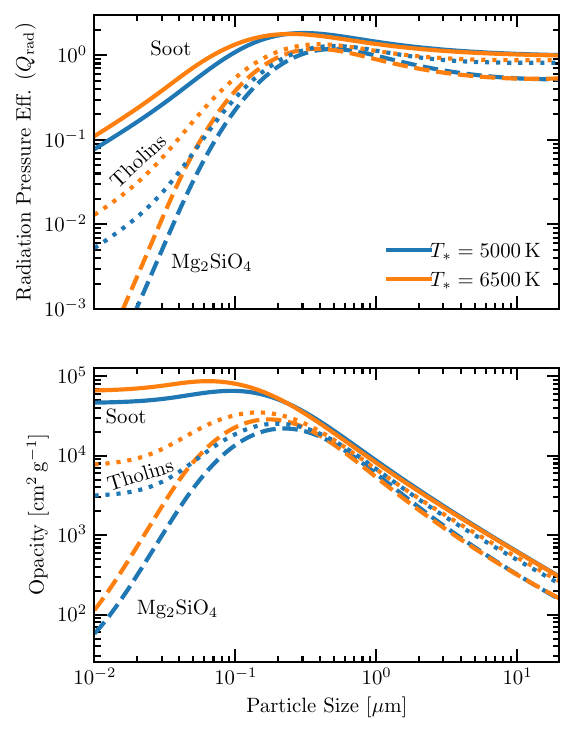}
    \caption{(Top) Radiation pressure efficiency, $Q_{\rm rad}$, as a function of particle size and stellar effective temperature, plotted for soot (solid), tholins (dotted) and silicate (Mg$_2$SiO$_4$, dashed) grains. (Bottom) Radiation pressure opacity.  Since $\kappa_{\rm rad}\propto Q_{\rm rad}/a$, the strength of the radiation pressure peaks at slightly smaller particle sizes than the efficiency.  Thus, we expect radiation pressure to operate most effectively on particle sizes around $0.1-1~\mu$m in size. }
    \label{fig:efficiency}
\end{figure}

Due to momentum conservation, the stellar light absorbed and scattered by particles imparts a force on them. The magnitude of the acceleration imparted on the aerosol particles due to radiation pressure is given by (neglecting relativistic corrections):
\begin{equation}
    a_{\rm rad} = \kappa_{\rm rad} \frac{F_*}{c}\,,
\end{equation}
where $\kappa_{\rm rad}\equiv \sigma_{\rm rad}/m_d$ is the radiation pressure opacity on an aerosol particle with mass $m_d$ and cross-section $\sigma_{\rm rad}$, $F_*$ is the {\rc unattenuated stellar bolometric flux at the planet's location} (under the assumption the radiation pressure is dominated by the star's bolometric emission) and $c$ is the speed of light. As typically done, comparing the magnitude of radiation pressure to other relevant forces is convenient. The other dominant force of particles in atmospheres is the planet's gravity. Therefore, defining $\beta_p$ as the ratio of the magnitude of the force due to radiation pressure due to the magnitude of the force due to gravity (c.f. \citealt{Wilck1996}), we find:
\begin{equation}
    \beta_p = \frac{\kappa_{\rm rad} F_* R_p^2}{G M_p c}\,, \label{eqn:beta_p_defn}
\end{equation}
with $R_p$ and $M_p$ the planet's radius and mass, respectively. In order to get a sense of the importance of radiation pressure compared to planetary gravity, we can write $\kappa_{\rm rad}=3Q_{\rm rad}/(4\rho_{\rm in}a)$ (appropriate for spherical particles of radius $a$) where $Q_{\rm rad}$ is the radiation pressure efficiency defined such that $\sigma_{\rm rad} = Q_{\rm rad}\pi a^2$, and $\rho_{\rm in}$ is the internal density of the particles. This allows us to estimate $\beta_p$ as (with $\rho_{\rm in}=1~{\rm g~cm^{-3}}$ {\rc following \citealt{Lavvas2011}, similar to values found experimentally by} \citealt{Horst2013}):
\begin{equation}
    \beta_p \approx 5 ~Q_{\rm rad} \left(\frac{a}{0.1~{\rm \mu m} }\right)^{-1}\left(\frac{T_{\rm eq}}{1500~K}\right)^4 \left(\frac{R_p}{1.5~{\rm R_J}}\right)^2\left(\frac{M_p}{0.5 ~{\rm M_J}}\right)^{-1} 
\end{equation}
where we have written the bolometric flux in terms of the planet's black-body equilibrium temperature ($T_{\rm eq}$). In the case of geometric optics, $Q_{\rm rad} \sim 1$, thus radiation pressure is important for the dynamics of particles in the atmospheres of close-in giant planets. In reality, $Q_{\rm rad}$ varies with particle composition, size and the star's spectrum. Since $Q_{\rm rad}$ falls off when the particle size is smaller than the typical wavelength of the radiation, the radiation pressure acceleration (which scales with $\kappa_{\rm rad} \propto Q_{\rm rad}/a$) will be maximised at a particle size of order $0.1-1.0$~$\mu$m (for stellar light sources, see Figure~\ref{fig:efficiency}). These sizes are around those that can dominate the opacity in observations of exoplanet atmospheres \citep{Powell2018}. Thus, for highly-irradiated planets, we expect radiation pressure {\rc might} be important in controlling the observables of exoplanet atmospheres.

{\rc In Equation~\ref{eqn:beta_p_defn} we have compared radiation pressure to gravity. However, vertical mixing is believed to be important for aerosol transport \citep[e.g.][]{Kawashima2019}, potentially explaining the inflated radii of ``superpuffs'' \citep[e.g.][]{Gao2020_Zhang} or the steep optical scattering slopes seen in some planets \citep[e.g.][]{Ohno2020}. Vertical mixing is typically approximated through a diffusion law, with a diffusion constant, $K_{zz}$, where one can define a typical diffusive velocity as:
\begin{equation}
    u_{\rm diff}\sim \frac{K_{zz}}{H}
\end{equation}
where $H$ is the atmospheric pressure scale-height. The particle distribution is dominated by vertical mixing, rather than radiation-pressure-enhanced settling if this diffusive velocity exceeds the settling velocity. Adopting the terminal velocity approximation and the ballistic approximation for aerosol-gas coupling relevant for the upper regions of exoplanet atmospheres (see Section~\ref{sec:anal}, \citealt{Ormel2019}) in the limit that radiation pressure dominates over gravity, the settling velocity is approximately:
\begin{equation}
    v_{\rm set} \sim Q_{\rm rad}\frac{F_*c_s}{Pc}
\end{equation}
with $c_s$ the isothermal sound-speed of the gas and $P$ the gas' thermal pressure. Thus, we find that radiation pressure dominates over vertical mixing when:
\begin{eqnarray}
    K_{zz} &\lesssim& 3\times10^{9}~{\rm cm}^2~{\rm s^{-1}}\frac{Q_{\rm rad}}{\mu^{3/2}} \left(\frac{P}{1~{\rm mbar}}\right)^{-1} \left(\frac{T_{\rm eq}}{1500~{\rm K}}\right)^{11/2} \nonumber \\ &\times&\left(\frac{R_p}{1.5~{\rm R_J}}\right)^2\left(\frac{M_p}{0.5 ~{\rm M_J}}\right)^{-1}
\end{eqnarray}
where $\mu$ is the mean-molecular weight of the atmospheric gas. The value of $K_{zz}$ is highly uncertain on exoplanets \citep[e.g.][]{Parmentier2013}.  {\rcc This critical value of $K_{zz}$ straddles those predicted for hot exoplanets. For example, \citet[][]{Komacek2019} found values in the range $10^9-3\times10^{10}$~cm$^2$~s$^{-1}$ at 1~mbar for a HD 209458b-like planet with equilibrium temperatures in the range 1000-2000~K. \citet{Parmentier2013} measured a value around $10^{10}$~cm$^2$~s$^{-1}$ for HD 209458~b at 1mbar, while \citet{Agundez2014} reported a value of $5\times10^8$~cm$^2$~s$^{-1}$ for HD 189733~b. Therefore, given the wide spread in reported values, there will likely be regions of parameter space where radiation pressure dominates over diffusion, and regions where diffusion dominates over settling/radiation pressure, particularly as the parameter space of highly irradiated, lower density Saturns/sub-Saturns has yet to be explored with simulations. Additionally, our critical $K_{\rm zz}$ increases as pressure decreases more rapidly than parametrisations in the literature \citep[e.g.][]{Moses2011,Parmentier2013,Agundez2014,Noti2024}. Thus, radiation pressure is likely to dominate over vertical mixing at low pressures controlling aerosol dynamics at these altitudes, potentially impacting observables, even if vertical mixing dominates at higher photospheric pressures. } 

}

\subsection{Comparison to Exoplanet Systems}

\begin{figure*}
    \centering
    \includegraphics[width=\textwidth]{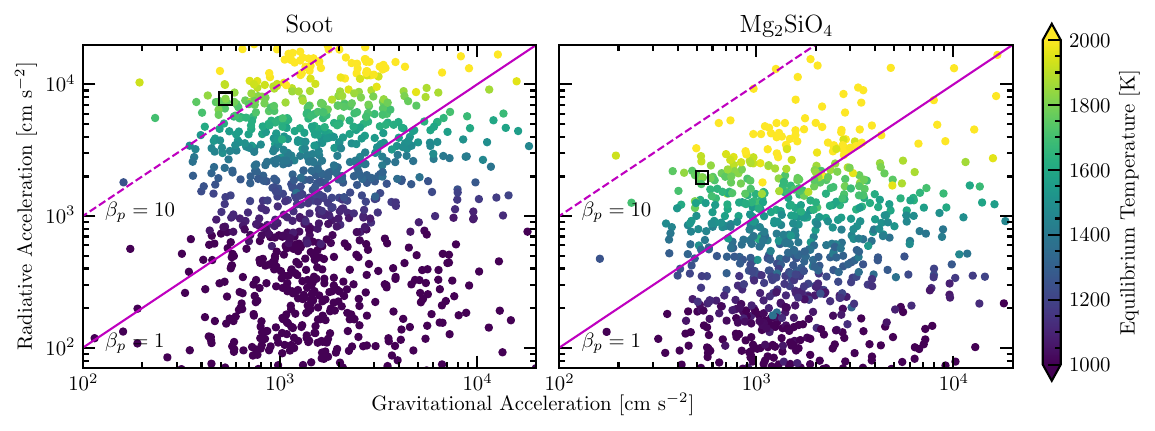}
    \caption{The population of exoplanets showing the radiative acceleration of a 0.1~$\mu$m aerosol particle (soot: left panel, silicates: right panel) in their atmosphere as a function of the planet's gravitational acceleration and equilibrium temperature. The square point highlights the planet HAT-P-65~b, whose parameters we will adopt in our later simulations. The lines indicate constant values of the ratio of the radiative acceleration to the planet's gravitational acceleration $\beta_p$. Tholin-like hazes have radiative acceleration strengths between soot and silicates. For the population of observed planets with measured mass and radii, $\beta_p$ exceeds 1 for 39\% and 17\% of the planets with soot and silicate particles, respectively. This rises to 52\% and 24\%, respectively, when one considers planets with likely primordial atmospheres (crudely defined using densities $<3~$g~cm$^{-3}$). }
    \label{fig:exoplanet_data}
\end{figure*}

To compute the strength of radiation pressure,  we use the optical constants of soot \citep{Lavvas2017}, tholins \citep{Khare1984} and silicates  (Mg$_2$SiO$_4$, \citealt{Jager2003}) to calculate the radiative efficiencies for uniform spherical particles. We use the modified version of the {\sc bhmie} code \citep{Bohren1998} provided by the {\sc dsharp opac} package \citep[e.g.][]{Birnstiel2018}. The frequency-dependent radiation pressure efficiency, $Q_{\rm rad}^\nu$, is then calculated as $Q^\nu_{\rm abs}+(1-g^\nu)Q^\nu_{\rm scat}$ \citep[e.g.][]{Burns1979,Bohren1998,Weingartner2001}, with $Q^\nu_{\rm abs}$ and  $Q^\nu_{\rm scat}$ the frequency-dependent absorption and scattering efficiencies respectively, and $g^\nu$ is the frequency-dependent asymmetry parameter. $Q_{\rm rad}$ is then averaged over a Planck spectrum at the star's effective temperature.  The optical constants and opacities are plotted in Figure~\ref{fig:efficiency}. {\rc Soot hazes are believed to dominate over tholins at high temperatures \citep{Lavvas2017}, which might be more applicable to planets where radiation pressure is important. In this work, we choose to be agnostic about the aerosol dependence on planetary properties to understand how the impact of radiation pressure varies with different aerosol radiative properties.} 

To understand the types of planets that are most strongly affected by radiation pressure, we use our calculated opacities to determine the radiative acceleration on 0.1~$\mu$m soot and silicate (Mg$_2$SiO$_4$) aerosol particles when exposed directly to the parent star's radiation (given Figure~\ref{fig:efficiency}, tholins sit in between these two). The result of this exercise is shown in Figure~\ref{fig:exoplanet_data}, where we compare the radiative acceleration on our aerosol particle to the planet's gravitational acceleration (evaluated at the transit radius) for the observed exoplanet population \footnote{Downloaded from the NASA exoplanet archive on 10th January 2025.}. We find hot exoplanets tend to have $\beta_p$ values up to $\sim$10-30, with HAT-P-67 b having the largest value. 

As radiative acceleration correlates strongly with the planet's equilibrium temperature, the highest radiative accelerations experienced by planets often have high equilibrium temperatures. In particular, we identify inflated, low-mass giant planets and ``sub-Saturns'' as the type of planets most likely to be affected. We also note that while sub-Neptune planets can have quite low gravities \citep[e.g.][]{JontofHutter2016}, the atmospheres of such planets are likely photoevaporated away at separations where radiation pressure would be important at ages of a few billion years \citep[e.g.][]{Owen2013,Rogers2021}. However, in Section~\ref{sec:discuss}, we discuss the effect of radiation pressure on aerosols in the atmospheres of young, low-mass planets. 

Unsurprisingly, favouring low-gravity planets with high-equilibrium temperatures for the role of radiation pressure strongly correlates with the potential observability of an exoplanet's spectrum. In transmission, the amplitude of a spectral feature is also largest for planets with high-equilibrium temperatures and low gravities \citep{Kempton2018}. Specifically, some of the inflated giant planets we identify in the extreme top-left of Figure~\ref{fig:exoplanet_data} have atmospheric scale heights $\gtrsim 1000$~km. Therefore, those planets for which it is easiest to observe their atmospheres in transmission are most likely to be affected by radiation pressure. To study the role of radiation pressure in these atmospheres, we take the mass and radius of the inflated, low-mass giant planet HAT-P 65b ($M_p=0.53$~M$_J$ and $R_p=1.89$~R$_J$, $T_{\rm eq}=1854~K$) along with its stellar properties as our archetypal planet.  {\rc HAT-P 65b has been observed by JWST (GO-3989, \citealt{Espinoza2023}), further motivating its choice as our archetypal planet.} 

\section{Hazy atmospheres}
In order to isolate the impact of radiation pressure on aerosol distributions, we choose to focus on ``haze'' particles. This allows us to ignore the impact of condensation and thermal feedback; however, we speculate about the impact of radiation pressure on cloud condensates in Section~\ref{sec:discuss}. By modelling haze particles in this initial work, we limit the growth of the particles to coagulation alone, allowing us to understand the impact of radiation pressure in a simpler system. 

\subsection{1D analytic sub-stellar point models}\label{sec:anal}
Before we move onto numerical simulations in Section~\ref{sec:numerics}, exploring the role of radiation pressure within an analytic framework is insightful. We assume that haze particles are produced at high altitudes with a constant flux ($\dot{\Sigma}_p$). Additionally, we ignore longitudinal transport for now, so we can consider a one-dimensional model at the planet's sub-stellar point (we will consider transport in our 2D models). At the sub-stellar point, radiation pressure and planetary gravity accelerate the particles in the same direction, hence the particles feel an acceleration of $(1+\beta_p)g$, where $g$ is the strength of the planet's gravitational acceleration.  

As the particles settle under the influence of gravity and radiation pressure, they grow, reaching a growth-settling equilibrium where the growth and settling timescale are equal. At the altitudes relevant for observations, the haze particles typically have gas Knudsen numbers (the ratio of the gas mean-free-path to particle size) much greater than unity, meaning that the ballistic approximation appropriately describes their dynamics \citep[e.g.][]{Rossow1978}. Thus, adopting the Epstein drag law, the stopping time is:
\begin{equation}
    t_{\rm stop} = \frac{a \rho_{\rm in}}{v_{\rm th}\rho_{\rm gas}}
\end{equation}
where $\rho_{\rm gas}$ is the gas density and {\rc $v_{\rm th}=\sqrt{8k_bT/\pi m_g}$ is the gas' thermal speed for a mean gas particle mass, $m_g$, and $k_b$ is Boltzmann's constant.} Adopting the short-friction time approximation yields the settling speed of the particles as follows:
\begin{equation}
    v_{z} = t_{\rm stop}\left(1+\beta_p\right)g
\end{equation}
For high altitudes and small particle sizes, relative velocities between the grains are dominated by ballistic ``Brownian'' motion, where the particles collision timescale is \citep[e.g.][]{Ormel2019}:
\begin{equation}
    t_{\rm col}^{-1} = 2\pi v_{\rm BM}a^2n_p \label{eqn:grow_simple1}
\end{equation}
where $v_{\rm BM}=\sqrt{16k_bT/\pi m_p}$, is the relative velocity for equal-mass particles with a mass $m_p$, in thermal equilibrium with gas at a temperature $T$, and $n_p$ is the particle number density. Assuming that all collisions result in growth, we may equate the collision timescale to the settling timescale $\sim H/v_z$, with $H$ the gas pressure scale height. Equating these two timescales, along with mass conservation ($\rho_p v_z = \dot{\Sigma}_p$, where $\rho_p$ is the particle mass density), we can obtain the growth-settling equilibrium, allowing us to write the typical particle size as:
\begin{equation}
    a = \left[\frac{24\dot{\Sigma}_pP^2\sqrt{3k_b T}}{\pi^2 \rho_{\rm in}^{7/2}\left(1+\beta_p\right)^2g^3}\right]^{2/9} \label{eqn:size_arad}
\end{equation}
and the particle {\rc mass} concentration as:
\begin{equation}
    X_p = \frac{\rho_p}{\rho_{\rm gas}} =   \left[\frac{8^{5/2}\left(k_bT\right)^{7/2}\dot{\Sigma}_p^{7}}{3^{3}\pi^{1/2}m_g^{9/2}\left(1+\beta_p\right)^{5}g^{3} \rho_{\rm in}^2 P^4}\right]^{1/9} \label{eqn:conc_arad}
\end{equation}
 Hence, we immediately see the impact of radiation pressure: radiation pressure causes the particles to settle faster, such that they end up smaller with a lower particle {\rc mass} concentration for the same production rate. {\rc In the following sections, we adopt $m_g=2.35m_h$, with $m_h$ the mass of the hydrogen atom, appropriate for solar metallicity molecular gas, $T=T_{\rm eq}$, and $\rho_{\rm in}=1$~g~cm$^{-3}$}.

\subsection{Impact on observations}
The observed properties of exoplanetary atmospheres will depend on the opacity profile of the aerosols. In the above analysis, we have assumed $\beta_p$ is a constant for simplicity; however, inspection of Figure~\ref{fig:efficiency} indicates this is not generally the case. In particular, the radiation pressure efficiency has a form where it is approximately constant and $\sim $ 1 for large grains ($a \gtrsim \lambda_*$, with $\lambda_*$ the typical wavelength of stellar photons). Whereas for small particles ($a \lesssim \lambda_*$), it has an approximately power-law dependence on particle size ($Q_{\rm pr} \propto (a/\lambda_*)^{1+b}$), where the index $b$ describes the Rayleigh limit. Therefore,  for small particles $\beta_p =\beta_0 (a/a_0)^{b}$ and $\beta_p = \beta_0(a/a_0)^{-1}$ for large particles, where $\beta_0$ is the strength of the radiation pressure for a particle size $a_0$. Inspection of Figure~\ref{fig:efficiency} indicates $b\sim$ 0, 0.6 \& 2.3 between 0.01 and 0.1 microns for soot, tholins and silicates, respectively. Thus, as discussed above, the typical impact of radiation pressure is that it increases as the particle size increases towards a size, $a \sim \lambda_*$ and then decreases for larger particles. Since aerosols in planetary atmospheres are typically smaller than $\lambda_*$ \citep[e.g.][]{Gao2021}, we expect radiation pressure to increase with particle size as they grow and settle.

Assuming that $\beta_p$ is large, so that $(1+\beta_p)\sim \beta_p$, allows us to consider the impact of radiation pressure on observations. We must find the opacity variation with height in the atmosphere (or pressure) to do this. The opacity of the aerosols ({\it per unit mass of gas}) is:
\begin{equation}
    \kappa_{g,p} = X_p \frac{3Q}{4\rho_{\rm in} a} \label{eqn:opac1}
\end{equation}
where $Q$ is the efficiency of the radiative process being considered (e.g. absorption, extinction, scattering, radiation pressure). In general, these efficiencies vary (e.g. $Q\ne Q_{\rm rad}$); however, to simplify our following general analysis, we assume that they have the same form \footnote{\rc Given $Q_{\rm rad}$ and $Q_{\rm ext}$ are simple order unity sums of $Q_{\rm abs}$ and $Q_{\rm sca}$ this means they will be within an order unity factor of each other. This assumption is relaxed in our numerical calculations.}. Including the dependence of $\beta_p$ on particle size, for small particles, in Equation~\ref{eqn:size_arad} yields the following dependency of the typical particle size and {\rc mass} concentration on pressure and radiation pressure for small particles:
\begin{equation}
    a\propto \dot{\Sigma}_p^{2/(9+4b)}\beta_0^{-4/(9+4b)}P^{4/(9+4b)}\label{eqn:size_scale}
    \end{equation}
    \begin{equation}
         X_p \propto  \dot{\Sigma}_p^{(7+2b)/(9+4b)}\beta_0^{-5/(9+4b)}P^{-4(1+b)/(9+4b)} \label{eqn:conc_scale}
    \end{equation}
    \begin{eqnarray}a(b=0.6)&\propto&\dot{\Sigma}_p^{0.18}\beta_0^{-0.35}P^{0.35}\\
     X_p (b=0.6) &\propto & \dot{\Sigma}_p^{0.72}\beta_0^{-0.44}P^{-0.56}
\end{eqnarray}
where we have also listed the scalings to 2~decimal places for tholin-like hazes ($b\approx0.6$) for reference. 
Without radiation pressure ($\beta_p = 0$), the above scalings become \citep[e.g.][]{Ohno2020}:
\begin{eqnarray}
    a&\propto& \dot{\Sigma}_p^{2/9}P^{4/9}\label{eqn:size_scale_noard}\\
    X_p &\propto & \dot{\Sigma}_p^{7/9}P^{-4/9} \label{eqn:conc_scale_no_arad}
\end{eqnarray}

{\rc We note that while the mass concentration decreases with increasing radiation pressure the number concentration ($C=n_p/n_g=X_pm_g/m_p$) actually increases, as the particles remain smaller:
\begin{equation}
    C \propto \beta_0^{7/(9+4b)}
\end{equation}
However, despite the higher number density of particles their smaller sizes mean that the coagulation rate ($\propto n_p a^2$, Equation~\ref{eqn:grow_simple1}) is suppressed at higher radiation pressures, explaining the smaller particles despite the higher number densities. 
}
Combining the dependencies on pressure from Equations~\ref{eqn:size_scale}~\&~\ref{eqn:conc_scale} with the definition of the opacity (Equation~\ref{eqn:opac1}) yields the following dependencies for the opacities on pressure and wavelength, with radiation pressure:
\begin{equation}
    \kappa_{g,p} \propto \dot{\Sigma}_p^{(7+4b)/(9+4b)} \beta_0^{-(5+4b)/(9+4b)}P^{-4/(9+4b)}\lambda^{-(1+b)}\end{equation}\begin{equation}
    \kappa_{g,p}(b=0.6) \propto  \dot{\Sigma}_p^{0.82}\beta_0^{-0.65}P^{-0.35}\lambda^{-1.6}\label{eqn:opac_rad}
\end{equation}
and, without radiation pressure:
\begin{eqnarray}
    \kappa_{g,p} &\propto& \dot{\Sigma}_p^{(7+2b)/9} P^{4(b-1)/9}\lambda^{-(1+b)}\\
    \kappa_{g,p}(b=0.6) &\propto& \dot{\Sigma}_p^{0.91} P^{-0.18}\lambda^{-1.6} \label{eqn:opac_wrad}
\end{eqnarray}
These results tell us that, in the case of $b>0$, as is common, the opacity has a weaker dependence on the production rate when radiation pressure dominates. Importantly, with radiation pressure, the opacity always decreases with increasing pressure; however, without radiation pressure, aerosols with Rayleigh indices $>1$ have opacities that increase with increasing pressure. 

The pressure of the photosphere ($P_{\rm ph}$) to incident stellar light is given approximately by $(\kappa_{g,p} + \kappa_{g,*})\rho_{\rm gas} H \sim 1$, where $\kappa_{g,*}$ is the gas opacity to optical star-light and the pressure scale height $H \sim P_{\rm ph}/(\rho_{\rm gas} g)$, so that:
\begin{equation}
    \left(\kappa_{g,p} + \kappa_{g,*}\right) P_{\rm ph} \sim g \label{eqn:photo_p} \;\;.
\end{equation}
Assuming that the opacity is entirely set by the aerosols ($\kappa_{g,p}\gg \kappa_{g,*}$), we find the photospheric pressure's dependence on wavelength, {\rc for fixed $g$}, is:
\begin{eqnarray}
        P_{\rm ph} &\propto& \beta_0 \lambda^{(9+4b)(1+b)/(5+4b)} \dot{\Sigma}_p^{-(7+4b)/(5+4b)} \\
        P_{\rm ph} (b=0.6) &\propto& \beta_0 \lambda^{2.46}\, \dot{\Sigma}_p^{-1.27}
\end{eqnarray}
which, in the absence of radiation pressure, becomes:
\begin{eqnarray}
    P_{\rm ph} &\propto & \lambda^{9(1+b)/(5+4b)} \dot{\Sigma}_p^{-(7+2b)/(5+4b)} \\
    P_{\rm ph} (b=0.6) &\propto & \lambda^{1.95}\, \dot{\Sigma}_p^{-1.11} 
\end{eqnarray}

\subsubsection{Reflected light}
The amount of scattered light by the aerosols in the planet's atmosphere is approximately:
\begin{equation}
    \sim \pi R_p^2 \omega(a^{\rm ph}_\lambda)
\end{equation}
where {\rc $\omega=Q_{\rm sca}/Q_{\rm ext}$} is the single scattering albedo and $a^{\rm ph}_\lambda$ is the particle size at the photosphere for a wavelength $\lambda$. For small particles ($a\lesssim \lambda$), the single scattering albedo increases as the ratio of particle size to wavelength increases with a form $\omega\propto (a/\lambda)^d$.  Thus, to understand the impact of radiation pressure on a planet's reflective properties, we need to understand the impact of radiation pressure on the particle size at the photosphere and the single-scattering albedo. Ignoring the gas opacity (assuming that aerosols dominate extinction), the photospheric pressure is inversely proportional to the aerosol extinction opacity (Equation~\ref{eqn:photo_p}). Combining Equations~\ref{eqn:size_scale}-\ref{eqn:photo_p} yields the particle size at the photosphere to a wavelength $\lambda$ as:
\begin{eqnarray}
    a_\lambda^{\rm ph} &\propto& \dot{\Sigma}_p^{-2/(5+4b)}\lambda^{4(1+b)/(5+4b)} \label{eqn:a_phot}\\
    a_\lambda^{\rm ph}(b=0.6)& \propto& \dot{\Sigma}_p^{-0.27}\lambda^{0.86}
\end{eqnarray}
This result tells us that the size of the particles in the photosphere is not explicitly dependent on the strength of the radiation pressure. {\rc While this is a correct physical result, one cannot separate the relative strength of radiation pressure from other parameters in reality (e.g. $\beta_p$ and the planet's gravity, or radiative acceleration and the planet's equilibrium temperature). As such, these {\it implicit} correlations mean that the particle size at the photosphere does vary slightly with radiation pressure strength when these are included.} In the absence of radiation pressure, the particle size at the photosphere satisfies the same equation (\ref{eqn:a_phot}). However, since we have changed the particle size distribution and {\rc mass} concentration above the photosphere, this column will have slightly different reflective properties. To account for these effects, to include gas extinction, and to investigate our results in more detail, we use Equations~\ref{eqn:size_arad} and $\ref{eqn:conc_arad}$ to numerically solve for the particle size and optical depth as a function of altitude for tholin-like hazes (ignoring attenuation on the radiation pressure acceleration at this stage). 

\begin{figure*}
    \centering
    \includegraphics[width=0.495\textwidth]{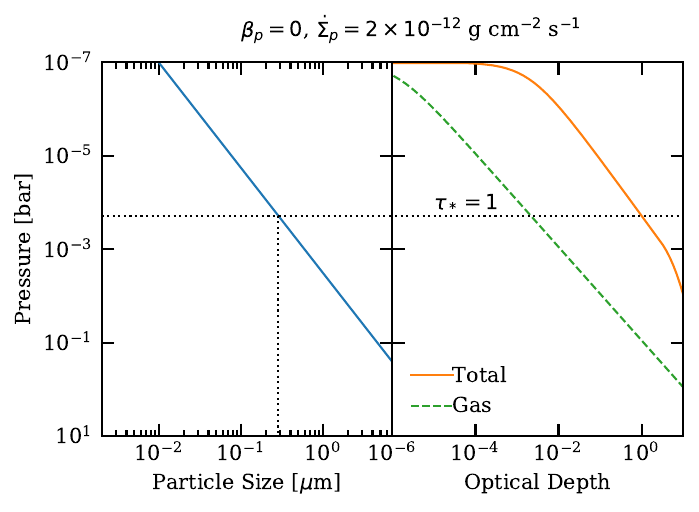}
    \includegraphics[width=0.495\textwidth]{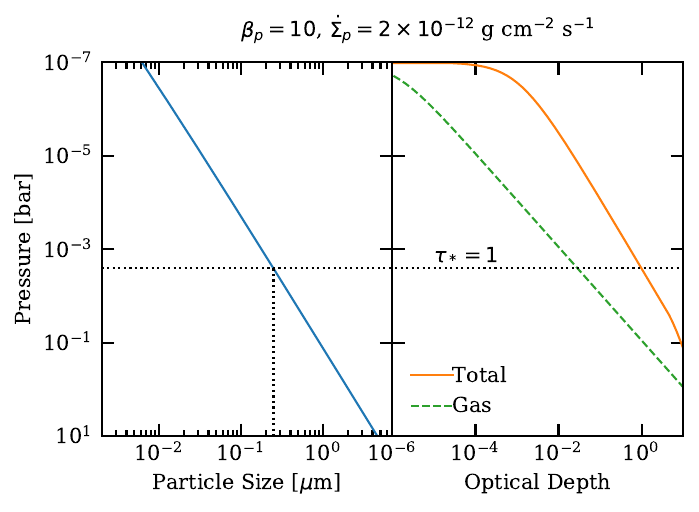}
    \includegraphics[width=0.495\textwidth]{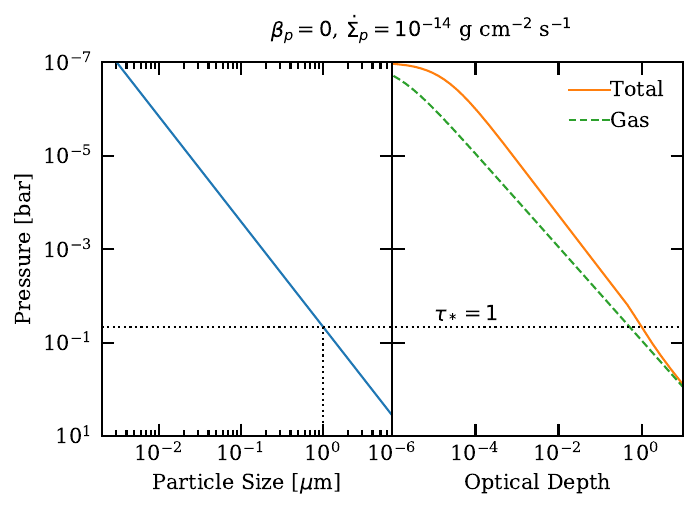}
    \includegraphics[width=0.495\textwidth]{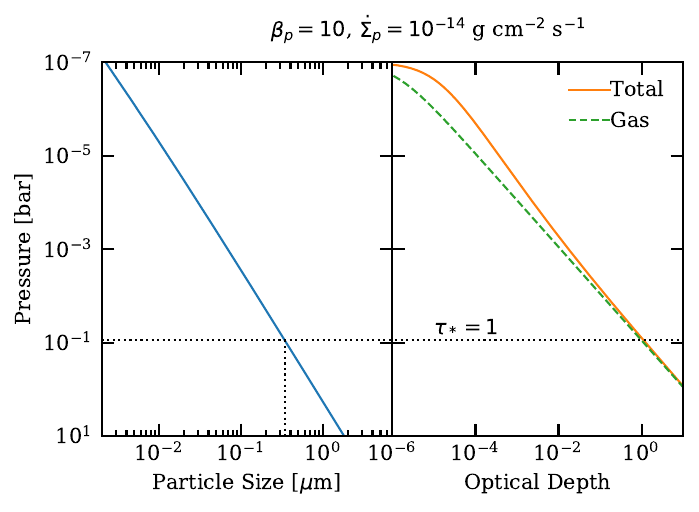}
    \caption{The optical depth, shown for the gas alone and for the gas and aerosols combined, and particle size as a function of altitude for no radiation pressure (left) and strong radiation pressure ($\beta_p(a=0.1~\mu{\rm m})=10$, right). For high production rates (top row), such that the aerosols control the position of the photosphere,  at high radiation pressure the lower particle density pushes the photosphere ($\tau_*=1$) deeper. However, since the particle size is smaller at a given pressure in the presence of strong radiation pressure, the particle size at the photosphere is weakly sensitive to the strength of the radiation pressure. For low production rates (bottom row), the photosphere position is set by gas opacity alone, thus due to the lower particle sizes in the case of high radiation pressure, higher radiation pressure results in smaller particles at the photosphere. As expected from Equation~\ref{eqn:size_scale}, radiation pressure results in the particle size varying less strongly with pressure.}\label{fig:compare_profiles}
\end{figure*}

Adopting the planetary parameters of HATP-65 b, we show the particle size and optical depth profiles as a function of pressure in Figure~\ref{fig:compare_profiles} for tholin-like hazes ($b=0.6$). We show a ``high'' production rate, where the optical depth is set exclusively by aerosols, and a ``low'' production rate, where the gas sets the photosphere. At high production rates, we see that even though radiation pressure pushes the photosphere to deeper pressure, the particle size at the photosphere is {\rc roughly} fixed as expected (Equation~\ref{eqn:a_phot}), {\rc where the weak variations come from the inclusion of gas opacity and implicit correlations between radiation pressure strength and planetary parameters}. Thus, for high production rates, we expect the planet's albedo to be weakly sensitive to radiation pressure. However, the photosphere remains at a fixed pressure (altitude) for low production rates. Thus, since radiation pressure produces smaller particles at a fixed altitude, we expect the albedo to be a strong function of the strength of radiation pressure.  

 Additionally, Figure~\ref{fig:compare_profiles} demonstrates that radiation pressure has changed the slope of the size (and density) distributions with altitude; in the presence of radiation pressure, the particle size distribution has a weaker dependence on pressure. This has important implications when we discuss transmission spectra. 

\begin{figure*}
    \centering
    \includegraphics[width=0.495\textwidth]{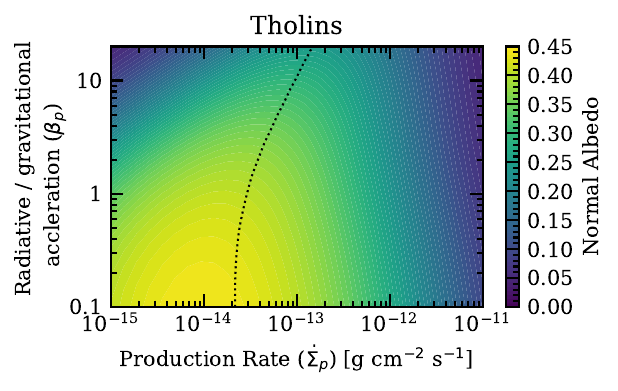}
    \includegraphics[width=0.495\textwidth]{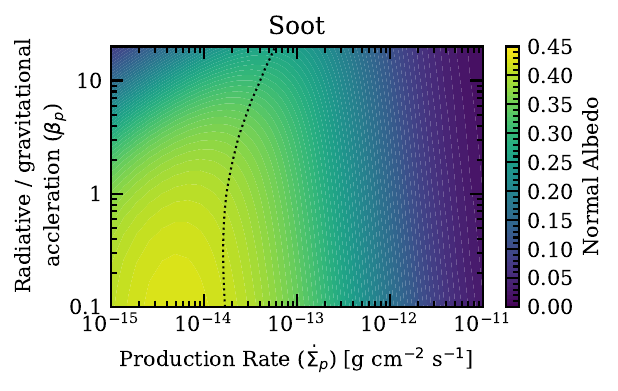}
    \caption{The fraction of incident stellar irradiation with a wavelength of 0.5$\mu$m scattered back at the sub-stellar point for different haze production rates and radiation pressure importance. The dotted line shows the line at which the gas and haze optical depths are equal at the photosphere, delineating between low and high production rates. The left panel shows tholin-like hazes, while the right panel shows soot-like hazes. For a given production rate, radiation pressure reduces the haze density.  Hence, at high radiation pressure, larger production rates are required to achieve high albedos. At the highest production rates, however, haze opacity sets the photospheric pressure at a height where the particle size only very weakly depends on radiation pressure, {\rc where this particle size decreases with increasing production rate}.  In this regime, higher production rates place the photosphere at higher altitudes where particles are smaller, leading to lower albedos. Taken together, high radiation pressure narrows the region of production rates where high albedos can be obtained.  }
    \label{fig:albedo}
\end{figure*}

 Using our 1D substellar models, we can calculate the ``normal albedo'' - i.e. the fraction of incident light reflected back by the sub-stellar point under the assumption of isotropic scattering (which is a good approximation for small particles). Adopting the single-scattering approximation and using the optical properties calculated in Section~\ref{sec:overview}, we estimate the fraction of reflected incident light, $R$, as:
 \begin{equation}\label{eqn:R}
     R = \int_0^\infty \omega(\tau)\exp\left(-\tau\right){\rm d}\tau
 \end{equation}
In this calculation, we assume the gas opacity to optical light is $4\times10^{-3}$~cm$^{2}$~g$^{-1}$ \citep{Guillot2010}. We adopted planet parameters for HAT-P-65 b again and varied the radiation pressure by changing the planet's distance from the host star. In Figure~\ref{fig:albedo}, we show the fraction of light reflected by the substellar region (normal albedo) as a function of the haze production rate and radiation pressure strength (where $\beta_p$ is scaled to a 0.1$\mu$m sized particle and $d=3$) for both tholin- and soot-like hazes. As expected, at low production rates, high radiation pressure leads to lower particle densities, smaller particles, and typically lower albedos.  At fixed radiation pressure, higher production rates generate more particles and, hence, higher albedos.  However, at very high production rates, the particle size at the photosphere depends only weakly on radiation pressure strength (particle size is independent of $\beta_p$ but depends weakly on correlated variations in equilibrium temperature as the planet's distance from its star changes). {\rc The highest production rates place the photosphere at higher altitudes where particles are smaller, leading to lower albedos.} Due to these combined effects, the range of production rates with high ($>0.3$) albedos is significantly smaller at high radiation pressure strengths---from 3-4 orders of magnitude for small radiation pressure strengths ($\beta_p\lesssim 0.1$) to only a factor of 2-3 for high radiation pressure strengths ($\beta_p \gtrsim 10$). Thus, radiation pressure potentially provides an additional mechanism to explain the variations in planetary albedos that, to date, cannot be correlated with standard planetary and stellar parameters \citep[see Section~\ref{sec:discuss}, and][]{Adams2022}. 

The amount of reflected light depends on the strength of the radiation pressure, which itself depends on several planetary parameters, as well as the production rate, which could also correlate with planetary parameters \citep[e.g.][]{Horst2018,Kawashima2019}. Thus, we choose to explore the reflected light signal in a more observationally accessible parameter space, with several choices for these potentially correlated properties.  

\begin{figure}
    \centering
    \includegraphics[width=\columnwidth]{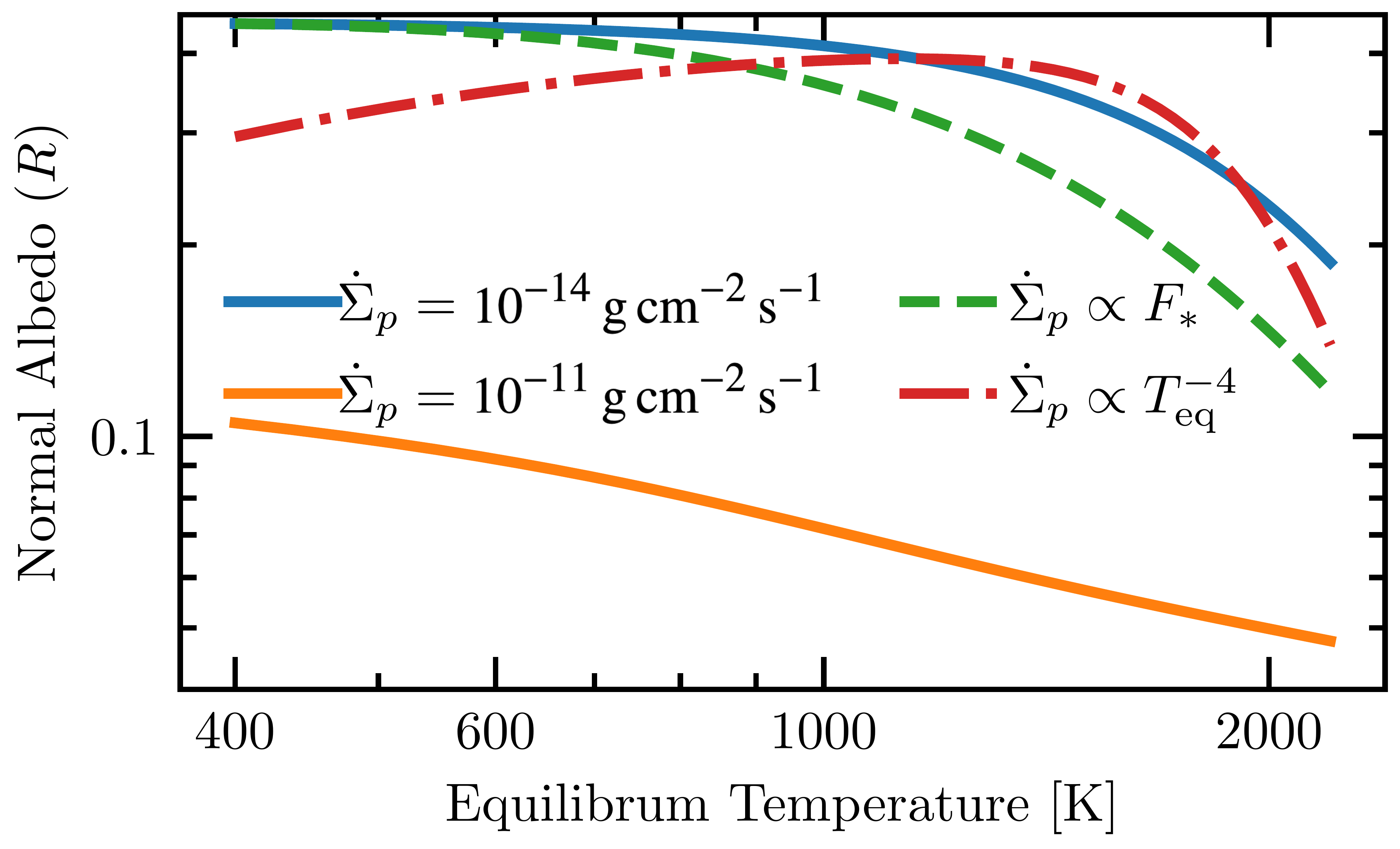}
    \includegraphics[width=\columnwidth]{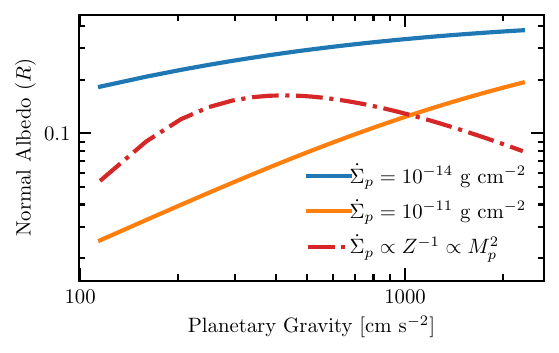}
    \caption{The fraction of reflected incident light from the sub-stellar point as a function of equilibrium temperature (top) and planetary gravity (bottom) for tholin-like hazes. We show two production rates of $10^{-14}$~g~cm$^{-2}$~s$^{-1}$ (blue) and $10^{-11}$~g~cm$^{-2}$~s$^{-1}$ (orange) as well as several production rates that implicitly correlate with planet properties. For equilibrium temperature, we correlate the production rate linearly with incident flux (green dashed), as the production rate is expected to increase with increasing UV flux. We also correlate it inversely with equilibrium temperature (red dot-dashed) as haze precursor molecules are expected to decrease with increasing equilibrium temperature. For planetary gravity, we correlate the production rate with planet mass, as increasing metallicity can provide more shielding to UV radiation, and metallicity is known to increase empirically with decreasing planet mass. }
    \label{fig:albedo_obs}
\end{figure}

In Figure~\ref{fig:albedo_obs} (top panel) we show the fraction of incident light reflected by our substellar point model as a function of the planet's equilibrium temperature (keeping the planet's bulk properties fixed to that of our HAT-P-65 b analogue), as in Figure \ref{fig:albedo}.  The equilibrium temperature is varied by adjusting the planet's distance from its host star, so that the radiation pressure strength varies accordingly.  In general, the trend shows a decrease in the albedo with increasing equilibrium temperature as the radiation pressure increases, resulting in lower particle {\rc mass} concentrations.  Two curves (blue and orange) correspond to slices through Figure \ref{fig:albedo} at fixed production rates.  However, given that the haze production rate may be implicitly correlated with the planetary parameters, we discuss two additional possibilities:
\begin{itemize}
    \item Higher UV fluxes are expected to increase the haze production rate \citep[e.g.][]{Horst2018,Kawashima2019}. The decreasing albedo with increasing equilibrium temperature trend is enhanced if the production rate increases with incident flux (e.g. $\dot{\Sigma}_p\propto F_*$; green dashed curve).
    \item Alternatively, since haze precursor molecules (such as methane) decrease in abundance as the equilibrium temperature increases, adopting a trend where $\dot{\Sigma}_p\propto T_{\rm eq}^{-4}$ - roughly consistent with the results of \citet{Kawashima2019}, we get a complex non-monotonic trend (red dot-dashed curve) where higher production rates yield higher albedos at low equilibrium temperatures before high radiation pressure produces low albedos at high equilibrium temperatures.
\end{itemize}

    Figure~\ref{fig:albedo_obs} (bottom panel) displays the planet's normal albedo as a function of the planet's gravity (fixing the equilibrium temperature to 1500~K and varying the mass from 0.1 to 2 M$_J$ with a radius of 1.5 R$_J$). When the planetary gravity is varied, we generally find an increasing albedo with increasing planetary gravity, as radiation pressure is less important at higher planetary gravities (blue and orange curves for fixed production rates).  However, production rates may also be implicitly correlated with gravity:  

\begin{itemize}
        \item  Higher metallicities can yield lower production rates due to shielding from molecules (like water) \citep{Kawashima2019}. Giant planet metallicities are known to empirically correlate with the planet's mass with an approximate form of $M_p \propto Z^{-1/2}$ \citep[e.g.][]{Thorngren2016}; if we suppose\footnote{Since due to shielding, the column of material exposed to UV irradiation will scale roughly as $1/Z$ as molecules dominate the opacity and metallicity.} that $\dot{\Sigma}_p \propto Z^{-1}$, then we find an implicit correlation of the form $\dot{\Sigma}_p\propto M_p^2$. Such an implicit correlation produces a non-monotonic albedo trend with planetary gravity (red dot-dashed curve), reversing the general trend at high gravities\footnote{We highlight that we do not actually suppose that $\dot{\Sigma}_p\propto M_p^2$, but rather use it as a possible implicit correlation that can cause complicated albedo trends.}.
    \end{itemize}
    In summary, because of the plethora of known (and unknown) implicit correlations that exist in the exoplanet population (e.g., hot Jupiter inflation), it is not surprising that the variation of planetary albedos is poorly understood \citep[e.g.][]{Adams2022}.

Given that the analytic model indicates that the particle size at the photosphere does not directly depend on radiation pressure, we choose not to explore reflection spectra here. Notwithstanding the difficulty of measuring it \citep[e.g.][]{Barstow2014}, there will be slight differences because radiation pressure can weaken the slope of the size distribution with altitude, meaning gas absorption plays a larger role. However, given these effects are minor and that the distributions of aerosols are likely to have a much larger influence on the temperature distribution \citep[e.g.][]{Morley2013,Juncher2017}, which in turn will impact any reflection spectra, we must wait until we understand the impact of radiation pressure on the temperature profile. 

\subsection{2D equatorial band models}
\begin{figure}
    \centering
    \includegraphics[width=\columnwidth, trim={0 5.5cm 31.1cm 0},clip]{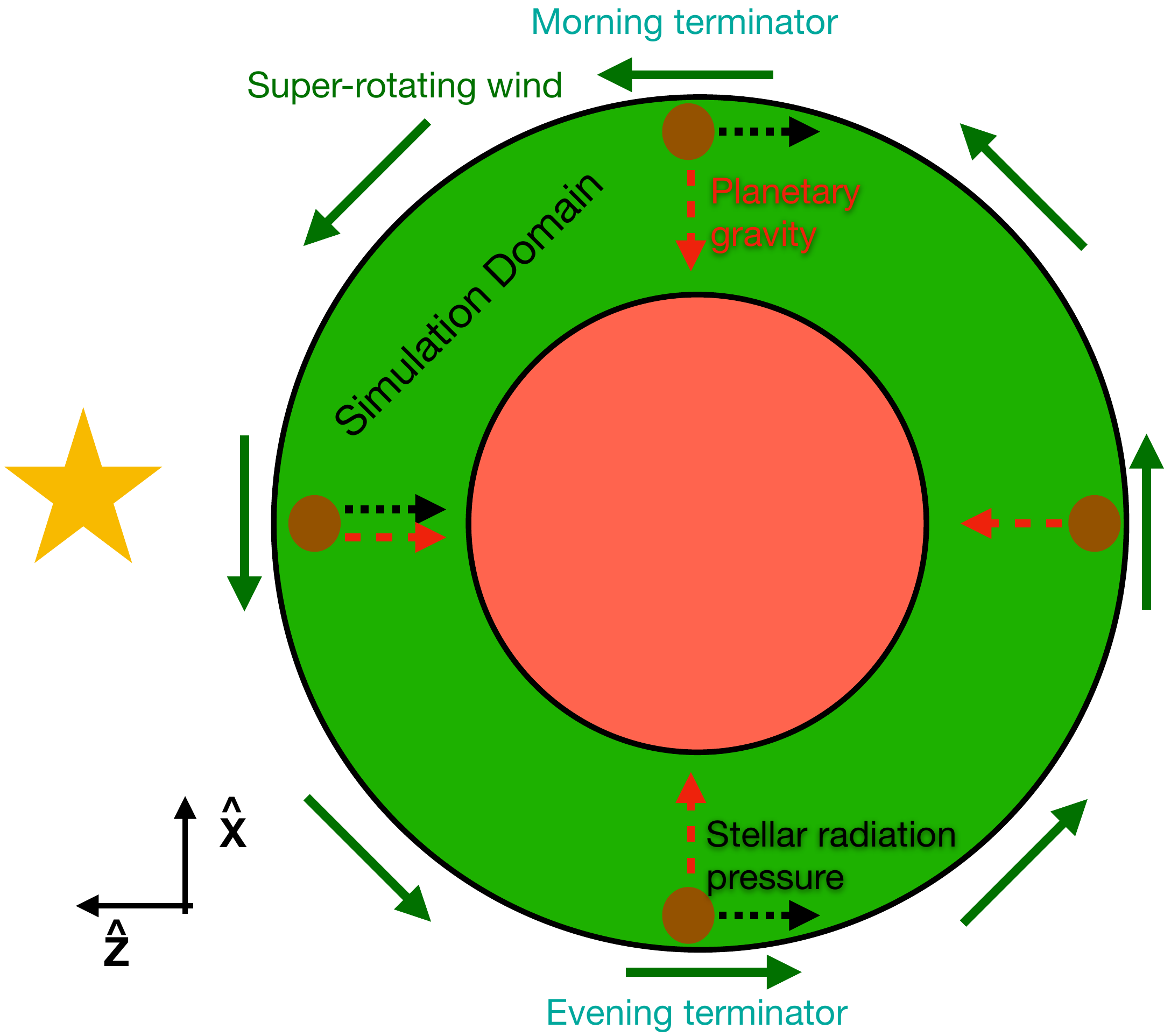}
    \caption{The schematic setup of our 2D equatorial band simulations, with our defined Cartesian coordinate system. The simulated band is shown in green. Aerosols with the directions of the gravitational force (red-dashed) and radiation pressure (black dotted) are shown at the substellar point, nightside and morning/evening terminator. The direction of the super-rotating wind that advects the aerosols is shown (green-solid).}
    \label{fig:2d_schematic}
\end{figure}

To begin the initial exploration of the role of radiation pressure on aerosol dynamics, we move beyond standard 1D models, which consider the particle size and density distribution in the vertical direction. Since radiation pressure imparts its force in the direction of the stellar irradiation, once we consider positions on the planet beyond the sub-stellar point, it is impossible to model the dynamics in 1D vertical slices.  Thus, we consider a simple 2D model centred on the planet's equatorial band \citep[similar to, ][]{Tsai2024}. This approach allows us to study the asymmetry between the morning and evening terminator: at the morning terminator, radiation pressure and advection from equatorial jets cause the particles to move in opposite directions, whereas at the evening terminator, radiation pressure and advection cause the particles to move in the same direction (see Figure~\ref{fig:2d_schematic}).

For simplicity, we consider a globally isothermal atmosphere which is in hydrostatic equilibrium (with a $1/r$ gravitational potential). Within this axisymmetric system, we then solve for the particle dynamics and size evolution. The continuity and momentum equations control the evolution of the particle dynamics:
\begin{eqnarray}
    \frac{\partial \rho_{p}}{\partial t} &=& - \nabla\cdot{\mathbf F} + S \label{eqn:2d_par}\\
    \frac{D{\mathbf u}}{Dt} &=& -\frac{{\mathbf u}-{\mathbf{u}_{\rm gas}}}{t_{\rm stop}} +{\mathbf g} + {\mathbf a}_{\rm rad} \label{eqn:mom}
\end{eqnarray} 
where ${\mathbf F}$ is the mass-flux of particles,  $S$ is the source of particles due to haze production (in mass per unit volume, per unit time), ${\mathbf u}$ is the particles' velocity, ${\mathbf u}_{\rm gas}$ the gas' velocity, ${\mathbf g}$ is the gravitational acceleration, ${\mathbf a}_{\rm rad}$ is the radiation pressure acceleration and $t_{\rm stop}$ is the stopping time. Finally, for the pressures and particle sizes we are most interested in {\rc we use the Epstein drag law as in Section~\ref{sec:anal}}. 

The mass-flux of particles arises due to both advection and diffusion, such that:
\begin{equation}
    {\mathbf F} = \rho_p{\mathbf u} - K\rho_{\rm gas}\nabla\left(\frac{\rho_p}{\rho_{gas}}\right)
\end{equation}
For simplicity, here we consider a constant isotropic diffusion constant, $K$. However, given $H/R_p\ll 1$, it is of limited importance in the longitudinal direction. Finally, to incorporate coagulation into our model, we consider the Lagrangian evolution\footnote{The choice of evolving the particle size in this manner naturally fits in our numerical scheme (Section~\ref{sec:numerics}).} of a particle's representative size as:
\begin{equation}
    \frac{D a}{Dt} = \frac{a}{t_{\rm grow}}
\end{equation}
For simplicity, we do not consider a particle size distribution, rather we consider the evolution of the particle distribution if it were peaked at this representative size. For the growth time due to coagulation, we implement the model of \citet{Ormel2019}\footnote{Correcting their Equation~12 to appropriately transition from the ballistic to diffusive Brownian motion regime.}, which incorporates contributions arising from differential particle drift and both ballistic and diffusive Brownian motion. Finally, to incorporate the acceleration arising from radiation pressure, we write this acceleration as:
\begin{equation}
    {\mathbf a}_{\rm rad} = - \frac{\kappa_{\rm rad}F_*\exp\left(-\tau_*\right)}{c}\hat{\mathbf z} \label{eqn:arad}
\end{equation}
where $\tau_*$ is the optical depth to stellar irradiation evaluated as:
\begin{equation}
    \tau_* = \int_\infty^z \left(\rho_{\rm gas} \kappa_{g,*} + \kappa_p(a,T_*) \rho_p\right){\rm d}z \label{eqn:tau_star}
\end{equation}
{\rc As in Section~\ref{sec:anal} we choose set $\kappa_{g,*}$ to the representative constant value of $4\times 10^{-3}$~cm$^2$~g$^{-1}$}, roughly appropriate for the values in a solar metallicity atmosphere \citep{Guillot2010,Freedman2014}. $\kappa_{p}(a,T_*)$ is the Planck mean extinction opacity of the particles to stellar irradiation with a temperature of $T_*$. This optical depth is evaluated using ray-tracing. Our form of the radiation pressure acceleration (Equation~\ref{eqn:arad}) implicitly assumes the stellar irradiation is plane parallel when it reaches the planet, with a constant flux $F_*$.  

\subsection{Source of particles}
Perhaps one of the most uncertain aspects of aerosol formation is the initial formation of the species. Detailed microphysical models can be used to simulate the process \citep[e.g.][]{Gao2018}; however, such models are complicated and are likely to shroud the basic physics we are trying to understand here. Therefore, we do not attempt to model aerosols' actual nucleation. Instead, we choose to prescribe the production rates of particles in our atmosphere. We chose the form proposed by \citet{Ormel2019}, which is used successfully in other works \citep[e.g.][]{Ohno2020}. In this parameterisation, the production rate of aerosols is simply given by a log-normal of the form:
\begin{equation}
    S = \rho_{\rm gas}g\frac{\dot{\Sigma}_p}{\sigma_*P\sqrt{2\pi}}\exp\left[-\frac{1}{2\sigma_*}\left(\log\frac{P}{P_*}\right)^2\right] \label{eqn:haze_produce}
\end{equation}
In this work we take $P_*=1~\mu$Bar and $\sigma_*=0.5$, where particles are inserted with a size of $10^{-3}~\mu$m. The 2D nature of our problems introduces two additional complications not present in previous works. Firstly, radiation pressure can transport grains to different longitudes. Thus, it is not {\it a priori} obvious what ``representative'' particle size to assign to a given position where particles are generated. If we consider the two limiting cases: either the newly inserted small particles coagulate amongst themselves, or they are swept up by the larger particles already present. Thus, when new particles are inserted, we determine which of the two will happen at a given location. In the case of the former, we do a mass-weighted average to determine the typical particle size; in the case of the latter, we compute the particle size the already present particles would grow to over the time step. This method also requires we force the time-step to ensure the timestep compared to the growth time is not so large that particles would grow significantly over this process. Thus, we limit the timestep to be, at maximum, $10\%$ of the growth timescale.  

\begin{figure}
    \centering
    \includegraphics[width=\columnwidth, trim={0 0cm 18.8cm 0},clip]{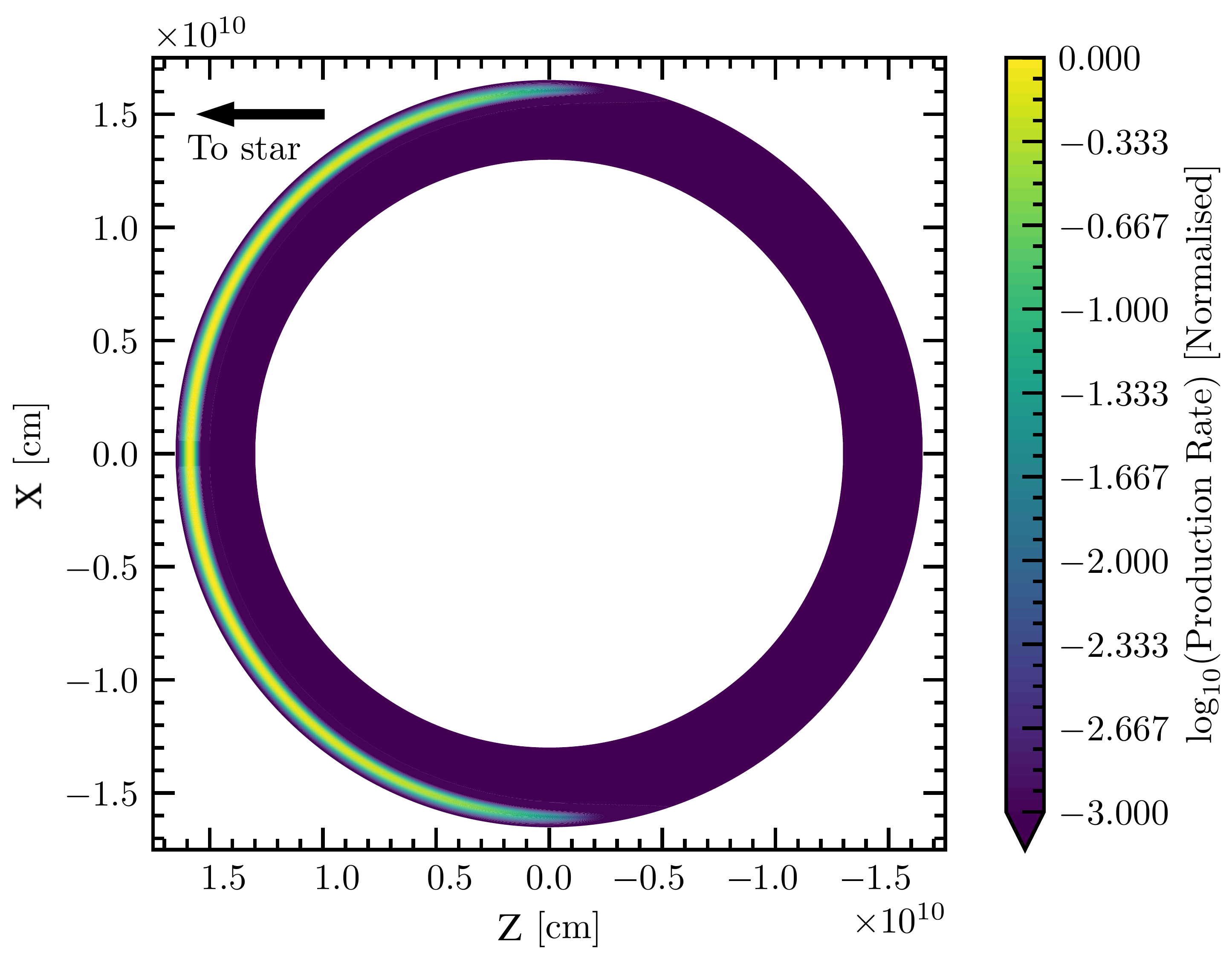}
    \caption{The profile of the haze production rate normalised to the maximum value. Our profile smoothly reduces the production rate towards the night-side.}
    \label{fig:production}
\end{figure}

Secondly, we need to specify the longitudinal dependence of our aerosol formation. As our initial exploration is on hazes, we assume they are generated by UV irradiation \citep[e.g.][]{Yu2021}. Thus, we track an additional optical depth $\tau_{\rm pro}$ that would cause the production rate to peak at $P_*$ (as explicitly adopted in Equation~\ref{eqn:haze_produce}) and then away from the sub-stellar point we reduce the production rate $S$ by a factor of $\exp(-\tau_{\rm pro})$. This approach means we smoothly shut off haze production on the night side. An example profile of the production rate for our standard simulation is shown in Figure~\ref{fig:production}.

\section{Numerical Simulations}\label{sec:numerics}

Our numerical code consists of two distinct components: an algorithm to evolve the particle density and size distribution and a ray-tracing scheme to evaluate the optical depth to stellar irradiation. To evolve the particle density and size distribution, we use an explicit scheme that is 1st order in time and 2nd order in space. The scheme is essentially an operator split {\sc zeus}-like scheme \citep[e.g.][]{Stone1992}, where we adopt a van-Leer limiter in the advection updates\footnote{We note in passing some more aggressive limiters introduce numerical instabilities in the presence of growth and radiation pressure.} and a CFL condition of 0.45. The particle distribution is updated on a staggered grid where scalar quantities (e.g. particle density, size, optical depth) are updated at cell centres, and vector quantities are updated at cell faces. Our spherical grid is logarithmically spaced in the radial direction and follows a power-law spacing in the angular direction, such that there is higher resolution at the terminators and lower resolution at the substellar point (and 180$^o$ from the substellar point on the night-side).
Further, we find that, for accurate advection (e.g., due to extremely strong radiation pressure) at the terminator, it is favourable to have square cells. Thus, we tune our resolution to have approximately square cells at the terminator. 

Experiments show that the terminal velocity approximation can break down at high altitudes for high radiation pressures. Thus, we update the momentum equation using the semi-implicit update of \citet{Rosotti2016}, which explicitly models the aerosol's acceleration up to the terminal velocity rather than adopting the short-friction timescale approximation.

\begin{figure}
    \centering
    \includegraphics[width=\columnwidth]{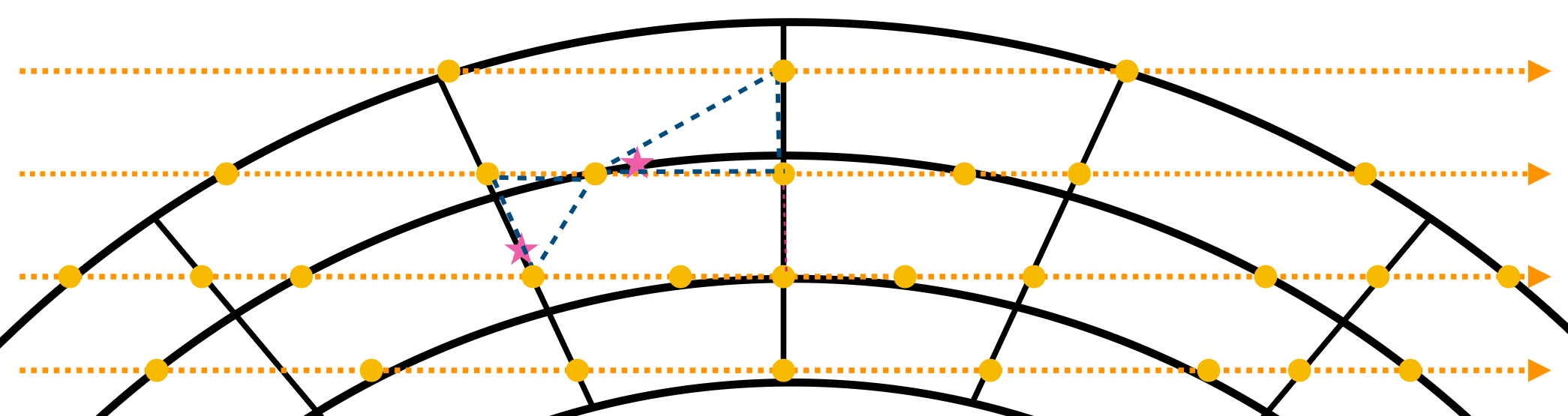}
    \caption{Schematic of our ray-tracing scheme. The cell faces are shown as solid black lines representing our spherical grid. Our plane parallel rays, shown as dotted orange lines, intercept the cell faces at the orange points where the optical depth is computed. Our aerosol dynamics scheme requires knowledge of the optical depth at the centre of the cell faces in both the radial and angular directions (magenta stars) to evaluate the acceleration due to radiation pressure. A Delaunay tessellation is constructed between the cell intercepts (blue dashed lines) in order to interpolate the optical depth values stored at cell intercepts to the cell face centres. }
    \label{fig:ray_tracing_example}
\end{figure}

To evaluate the radiation pressure, we need to evaluate the optical depth to the star (Equation~\ref{eqn:tau_star}). To do this, we construct a series of plane-parallel rays traversing our spherical grid. These rays then intercept the grid at various positions that do not necessarily align with the grid corners or cell centres. Thus, Equation~\ref{eqn:tau_star} is integrated using a first-order method between each cell intercept. This method then provides the optical depth at all the points at which the rays intercept the cell faces. The values of the optical depth at these intercepts are then used to compute the optical depths at the cell faces via linear interpolation on a Delaunay tessellation of these grid intercepts (see Figure~\ref{fig:ray_tracing_example}) for a schematic representation. We space the rays in a non-uniform manner, with a higher density of rays in the terminator region and a lower density around the sub-stellar point. At our standard running resolution, we can evaluate the optical depth with an accuracy of $\lesssim 1\%$ . Since ray-tracing is computationally time-consuming, we choose to update the optical depth every 20 time-steps of the density/size update. As we find our solutions evolve towards a steady state, this choice has no bearing on our final solutions. Furthermore, given the time-consuming nature of this on-the-fly radiative transfer, we do not interpolate our Planck mean opacities calculated in Section~\ref{sec:overview}. Instead, we fit the physically motivated functional form to the radiation pressure and extinction efficiencies separately:
\begin{eqnarray}
    Q &=& \frac{1}{1+\left[a(T_*/T_\odot)/\alpha_1\right]^{-\alpha_8}} \nonumber \\ &+& \left(\frac{\alpha_2}{\left\{\exp\left[\alpha_3T_\odot/(aT_*)\right]\right\}^{\alpha_4} + \left[a(T_*/T_\odot)/\alpha_5\right]^{\alpha_6}}\right)^{\alpha_7}
\end{eqnarray}
where $T_*$ is the stellar effective temperature and the eight $\alpha$ coefficients we fit for. Note how the first term reproduces the Rayleigh slope for small particles and the value of $Q\rightarrow 1$ for large particles. The second term provides a fit around the scattering bump when $a\sim \lambda_*$. We use {\sc scipy.optimize.curve\_fit} to fit this function to our calculated ones, finding it does an excellent job with errors of a few percent at maximum across the particle sizes of interest. 

This ray/grid structure is the basis for the calculations of our transmission spectra in Section~\ref{sec:spectra}; however, we use our calculated opacities as a function of the particle size and wavelength from section~\ref{sec:overview} for accuracy, where values are interpolated using {\sc scipy.interpolate.RectBivariateSpline}. Finally, our code is {\sc openmp} parallelised and accelerated using the {\sc numba} and {\sc numexpr} libraries \citep{Numba1,Numba}.

\subsection{Simulation setup}
We consider the evolution of particles in the atmospheres of a HATP-65 b-like planet. To understand the role of radiation pressure, we hold the planet and stellar properties constant but vary the value of the radiation pressure experienced by the planet. Specifically, we consider values of the stellar irradiation to yield values $\beta_p$ of 0 (e.g. no radiation pressure), 0.5, 1, 3, 6 and 10 for a 0.1~$\mu$m sized tholin-like or soot-like particles in HATP-65 b's atmosphere. We note that during this procedure, to isolate the role of radiation pressure, we hold its equilibrium temperature fixed to 1854~K for the observed values. We consider production rates from 10$^{-16}$ to $10^{-10}$~g~cm$^{2}$~s$^{-1}$ at the substellar point and diffusion constants from $10^5$ to $10^{10}$~cm$^2$~s$^{-1}$. This spans the full range from essentially particle-free to particle-dominated cases and from where diffusion is unimportant to where it dominates in the vertical direction. 

The pressure is set to 1 bar at a radius of $1.3\times10^{10}$~cm and extends up to a pressure of $\sim$0.1$\mu$bar at a radius of 1.65$\times10^{10}$~cm, we use 2000 cells in the angular direction, spaced symmetrically about the terminator in a fashion that scales as $\theta^2$ and 152 cells in the radial direction and 400 rays for the ray-tracing. This yields approximately ten cells per gas pressure scale height and twelve rays per gas pressure scale height in the terminator region. We confirm that our results are resolved by comparing them with simulations with a resolution of $1332\times100$ with 250 rays. 

\subsection{No zonal winds}\label{sec:no_zonal}
\begin{figure*}
    \centering
    \includegraphics[width=0.495\textwidth]{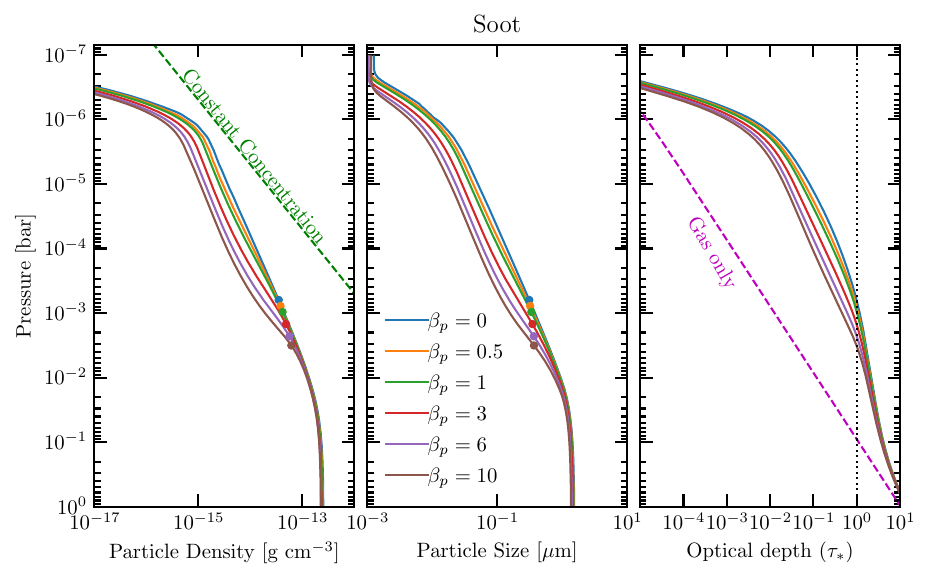}
    \includegraphics[width=0.495\textwidth]{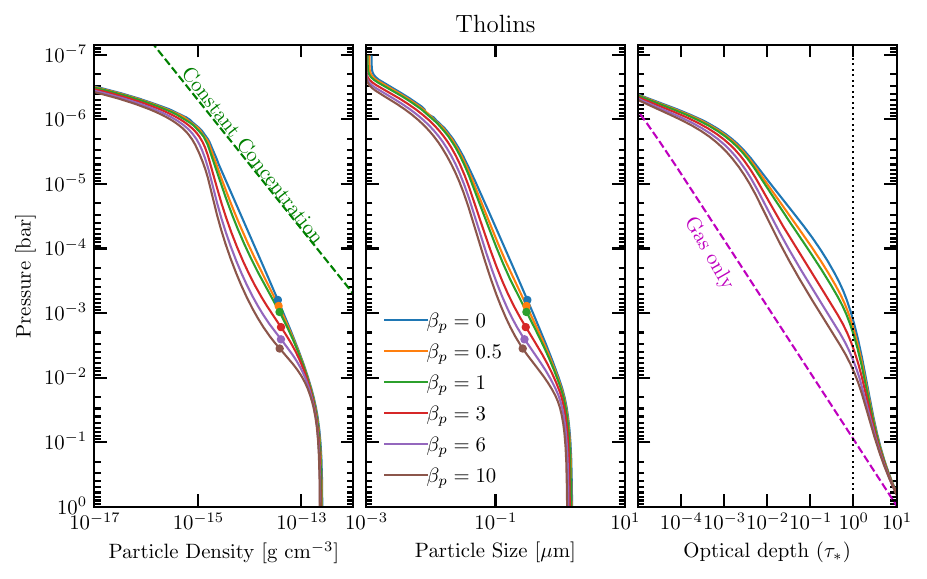}
    \caption{The evolution of the particle density, particle size and optical depth to stellar bolometric irradiation along the sub-stellar point for our HATP-67 b analogue. The lines are shown for different strengths of the radiation pressure, parameterised in terms of $\beta_p$ for a 0.1$\mu$m sized particle. The points in the particle density and particle size panels show the values at an value of the optical depth, $\tau_*=1$. The shown lines are for $\dot{\Sigma}_p=1.7\times10^{-13}$~g~cm$^{-2}$~s$^{-1}$ and $K=10^7$ cm$^2$~s$^{-1}$, with the left set for Soot-like particles and the right set row for tholin-like particles. As expected, higher radiation pressure leads to smaller particles and lower densities at a fixed altitude. {\rc The green dashed line shows the profile for a constant mass concentration and the dashed magenta line shows the optical depth contribution from only the gas. }}
    \label{fig:substellar_nowind}
\end{figure*}

To begin our exploration, we consider a case without a zonal wind to isolate the impact of radiation pressure on the hazes and compare it with our expectations from Section~\ref{sec:anal}. We show the evolution of the particle density and size, along with the optical depth to stellar bolometric irradiation along the substellar point in Figure~\ref{fig:substellar_nowind}.

These results indicate that radiation pressure lowers the particle density and size at a given altitude for a fixed production rate, pushing the photosphere to stellar irradiation deeper into the planet's atmosphere. For soot-like hazes, which have a Rayleigh index close to zero, the slope of the particle size distribution does not change; however, for tholin-like hazes, the increasing strength of radiation pressure with increasing particle size (while it remains small) results in a particle size distribution that varies less strongly with pressure than without radiation pressure, in agreement with our analytic expectations. We find that the general results do not depend on the strength of the production rate or the strength of diffusion (unless it is very high, in which case we reproduce the diffusion-dominated results of \citealt{Ohno2020}). Higher production rates result in larger particle concentrations and sizes, resulting in the photosphere to stellar bolometric radiation occurring higher in the atmosphere. 

\subsection{Inclusion of zonal winds}

\begin{figure}
    \centering
    \includegraphics[width=\columnwidth]{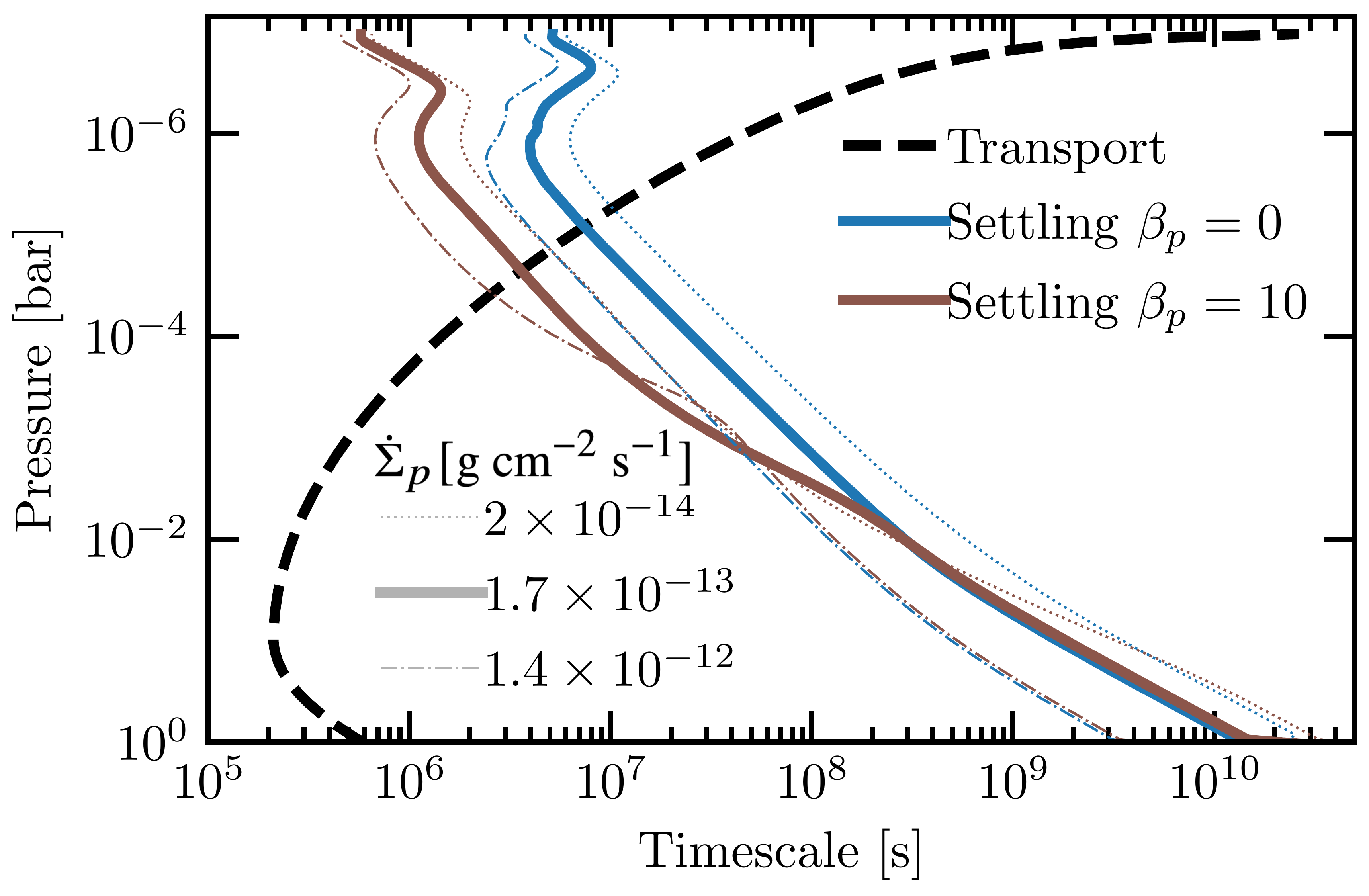}
    \caption{The timescale to transport material from the sub-stellar point to the evening terminator as a function of atmospheric pressure (black dashed) for a choice of $v_{\rm peak}=1$~km~s$^{-1}$. The local settling timescales ($H/u_z$) are shown for our HAT-P-65 b-like sub-stellar point soot haze models without zonal transport for different strengths of radiation pressure and production rate. Settling dominates at high altitudes, whereas transport dominates at lower altitudes, similar to the results of \citet{Tsai2024} for chemical evolution.  }
    \label{fig:advect_drift}
\end{figure}
It is well-established that many tidally locked giant planets have zonal winds that advect material from the day side to the evening terminator and the night side to the morning terminator \citep[e.g.][]{Showman2002}. To understand the role of zonal winds, we insert a longitudinally symmetric zonal wind profile that reproduces the mean zonal wind profiles measured in the equatorial band of GCMs. Zonal wind speeds typically peak near around $\sim 0.1$bar for giant planets. The phenomenological profile:
\begin{equation}
    v_{\rm zonal}(P)=v_{\rm peak}
    \begin{cases}
    \exp\left\{-\frac{\left[\log_{10}(P/P_0)\right]^2}{2\sigma_{\rm lp}^2}\right\},\quad{\rm when\,\,} P < P_0\\
    \exp\left\{-\frac{\left[\log_{10}(P/P_0)\right]^2}{2\sigma_{\rm hp}^2}\right\},\quad{\rm when\,\,} P > P_0
    \end{cases}
\end{equation}
with a choice of $P_0=0.09$~bar, $\sigma_{\rm lp} = 1.5$ and $\sigma_{\rm hp}=0.7$ matches the shape of the mean zonal wind profiles in \citet{Heng2011}. We emphasise that while this approach is not meant to be a detailed match, it elucidates the basic physics. We compare the local settling timescale (i.e., $H/u_z$) of our substellar point models without any zonal wind (those from Figure~\ref{fig:substellar_nowind}) to the advection timescale due to the zonal wind from the substellar point to the evening terminator (i.e. $\pi R / (2 v_{\rm zonal}$)) in Figure~\ref{fig:advect_drift}. This shows that at high altitudes, where the hazes are produced, low pressures and slower zonal winds mean that the settling timescale is faster than the advection timescale; however, at higher pressures, higher gas densities (resulting in slower settling times) and faster zonal wind speeds mean that the advection timescale is considerably smaller than the settling timescale. Namely, as also shown by \citet{Tsai2024} for chemical evolution, vertical transport dominates in the upper atmosphere, and longitudinal transport dominates deeper.  This transition means that the initial aerosol profiles established at high altitudes on the day side, which have lower densities and smaller particles as a result of radiation pressure,  will be advected to the terminators (and night side) by zonal winds. Thus, the transmission spectra will also be affected by the lower densities and smaller particle sizes, especially as for nominal values the region where the timescales become comparable is in the vicinity of the pressures probed in transmission \citep[e.g.][]{Fortney2005}. 
\begin{figure*}
    \centering
    \includegraphics[width=0.33\textwidth]{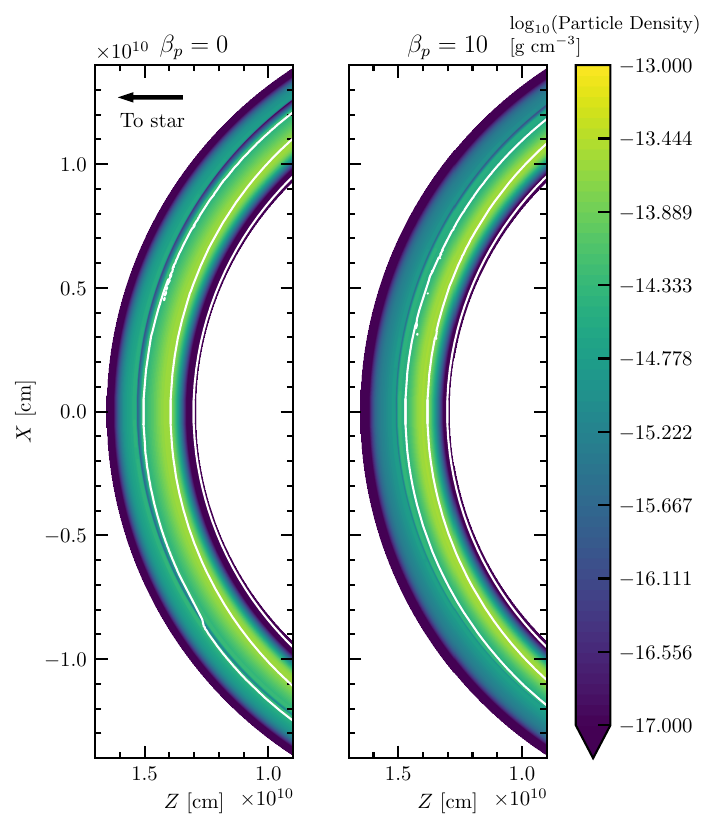}
    \includegraphics[width=0.33\textwidth]{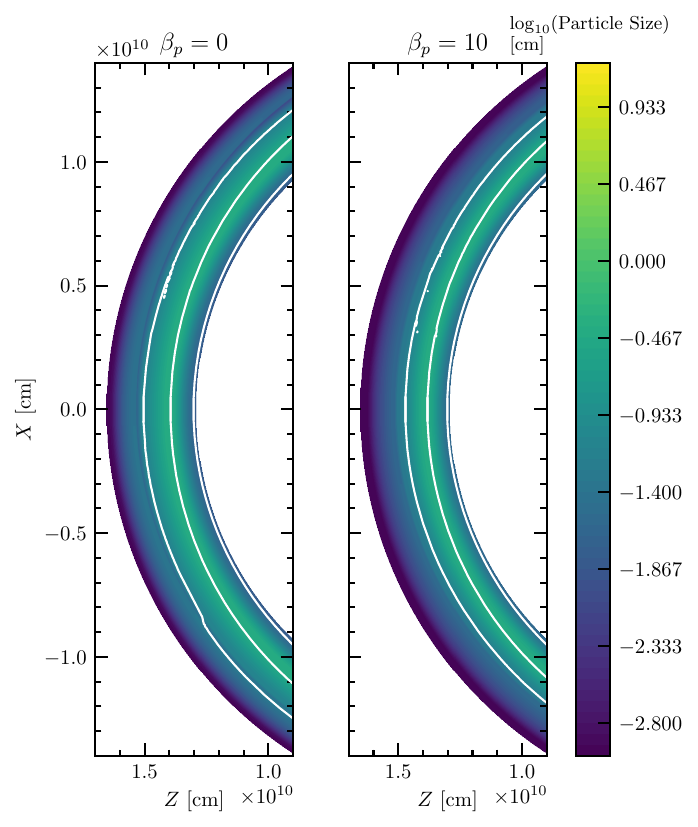}
    \includegraphics[width=0.33\textwidth]{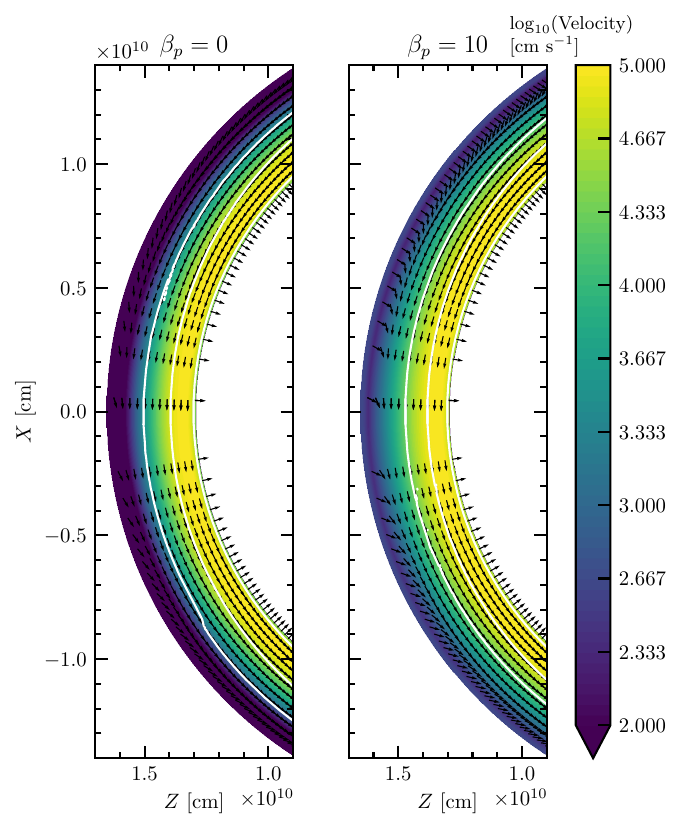}
    \caption{The density (left), particle size (middle) and velocity structure (right) on the day-side for simulations excluding radiation pressure (left) and including radiation pressure with a value of $\beta_p=10$ (right). The production rate is $\dot{\Sigma}_p=10^{-13}$~g~cm$^{-2}$~s$^{-1}$ and the diffusivity is $K=10^7$ cm$^{2}$~s$^{-1}$.  The white contours show the position of $\tau_*=$ 0.1, 1.0 and 10. The vectors in the right-hand panel are normalised such that the direction of the velocity is represented; the colour map shows its magnitude. Radiation pressure produces lower densities, smaller particles and a deeper photosphere. }
    \label{fig:compare_substellar}
\end{figure*}

\begin{figure*}
    \centering
    \includegraphics[width=0.9\textwidth]{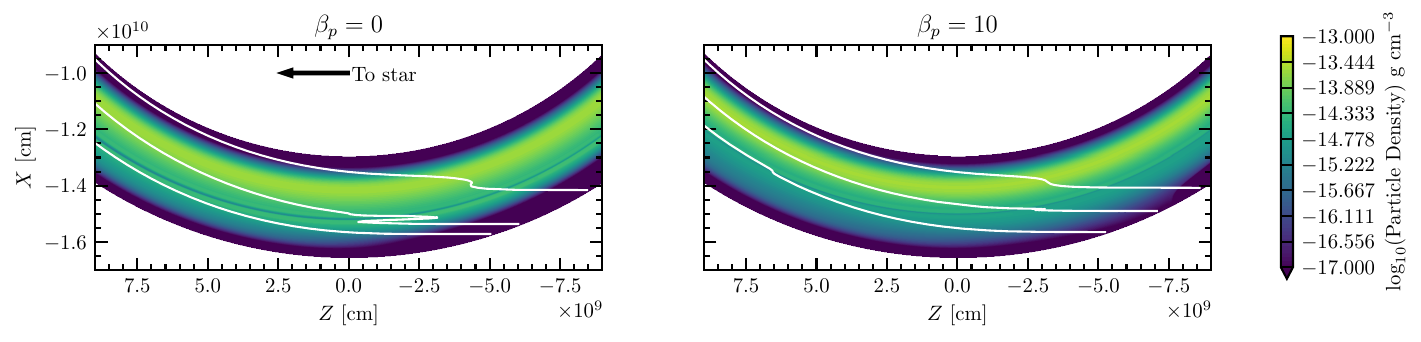}
    \includegraphics[width=0.9\textwidth]{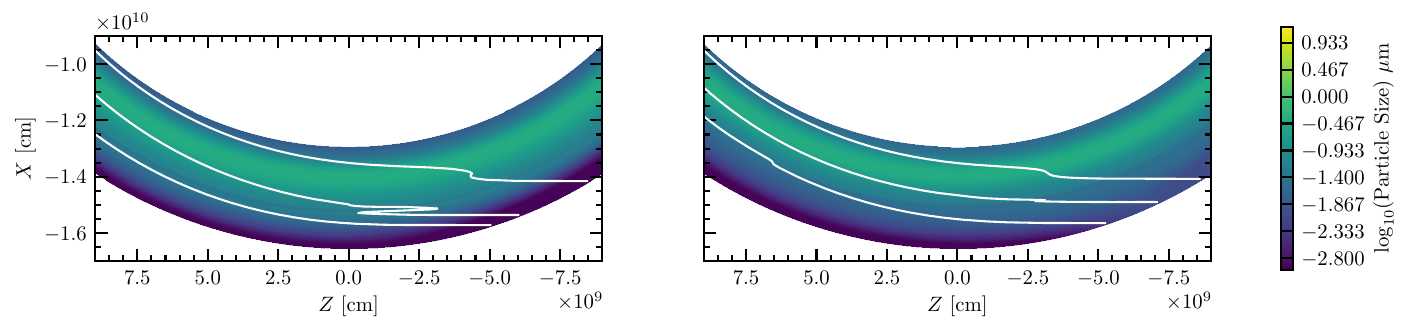}
    \includegraphics[width=0.9\textwidth]{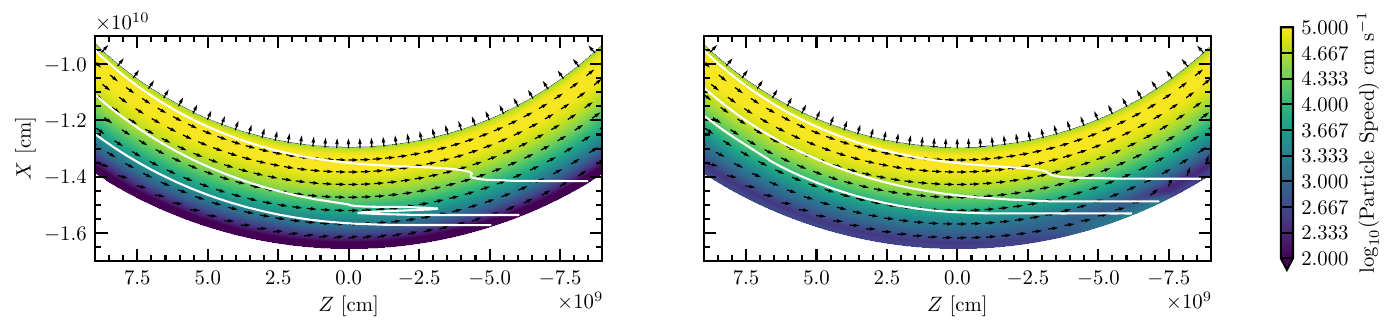}
    \includegraphics[width=0.9\textwidth]{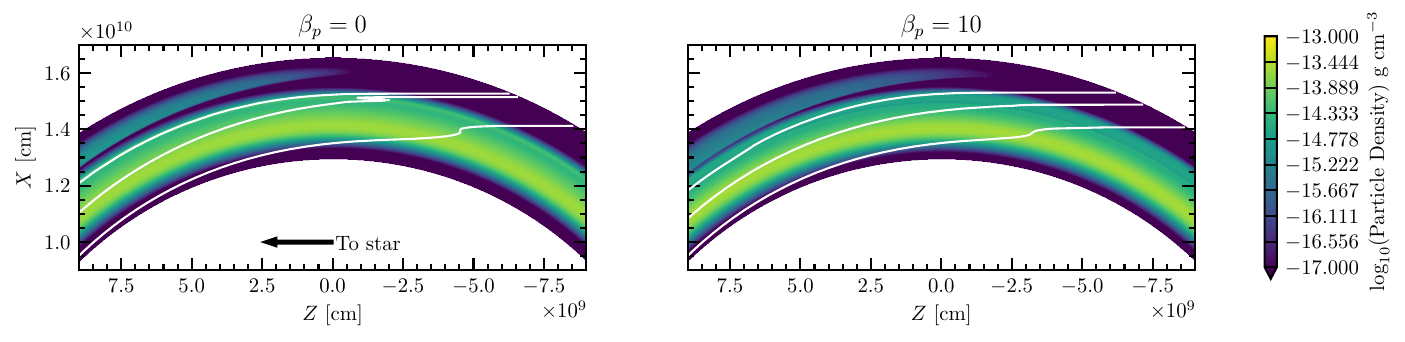}
    \includegraphics[width=0.9\textwidth]{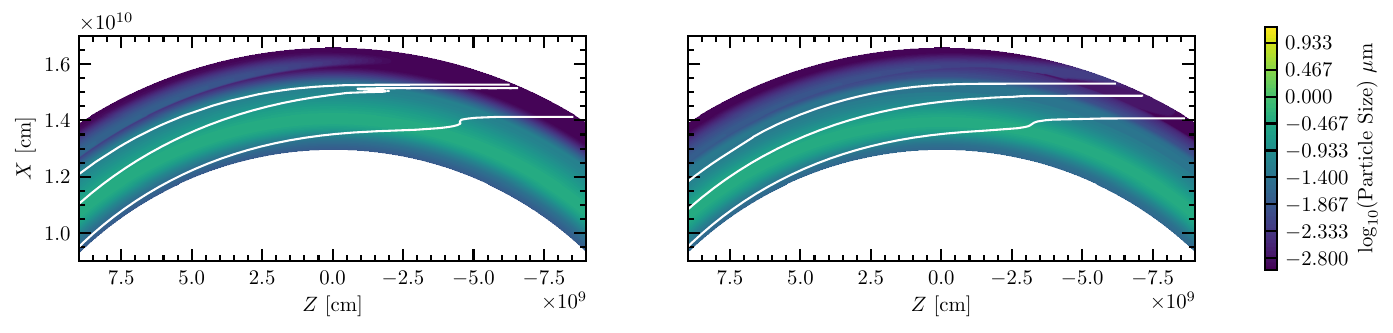}
    \includegraphics[width=0.9\textwidth]{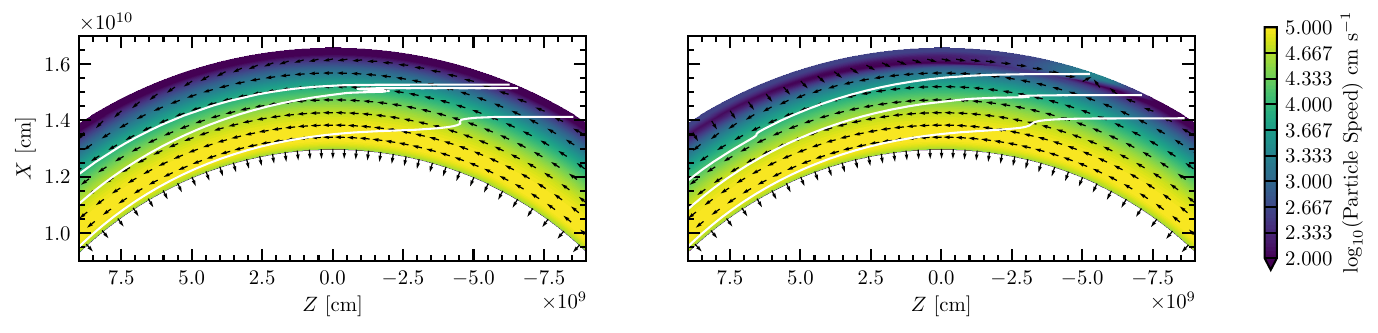}
    \caption{Similar to Figure~\ref{fig:compare_substellar}, but shown for the evening terminator (top three panels) and the morning terminator (bottom three panels), with the geometry as shown in Figure~\ref{fig:2d_schematic}. In the case of radiation pressure (right), the optical depth in the terminators is more slowly varying with height due to the shallower density and particle size gradients, which produce steeper scattering slopes (Section~\ref{sec:spectra}). }
    \label{fig:compare_term}
\end{figure*}
We now repeat our simulations from the previous section (\ref{sec:no_zonal}), but now include a zonal wind with $v_{\rm peak}=1$~km~s$^{-1}$. The zonal wind is included in our simulations by setting the gas velocity.  We show the profiles of the soot simulations without radiation pressure and with $\beta_p=10$ at the substellar point (Figure~\ref{fig:compare_substellar}), evening terminator (Figure~\ref{fig:compare_term}, top) and the morning terminator (Figure~\ref{fig:compare_term}, bottom). As expected, radiation pressure produces lower particle {\rc mass} concentrations and smaller particle sizes in all regions of the atmosphere. On the day side, this results in a deeper photosphere, while in the terminators, the shallower increase in density with increasing pressure results in lines of constant optical depth being further spread apart. This means that different wavelengths will have their respective photospheres at altitudes that are further apart, particularly at short wavelengths where Mie scattering-like slopes dominate. Thus, we expect radiation pressure will produce ``steeper'' scattering slopes in the optical.   The ``zig-zag'' features are caused by the zonal wind wrapping around material from the day-side {\rc that has settled while on the night-side}, meeting material settling from above, and it roughly occurs at the pressures where the transport and settling timescales match (Figure~\ref{fig:advect_drift}).  
\begin{figure}
    \centering
    \includegraphics[width=\columnwidth]{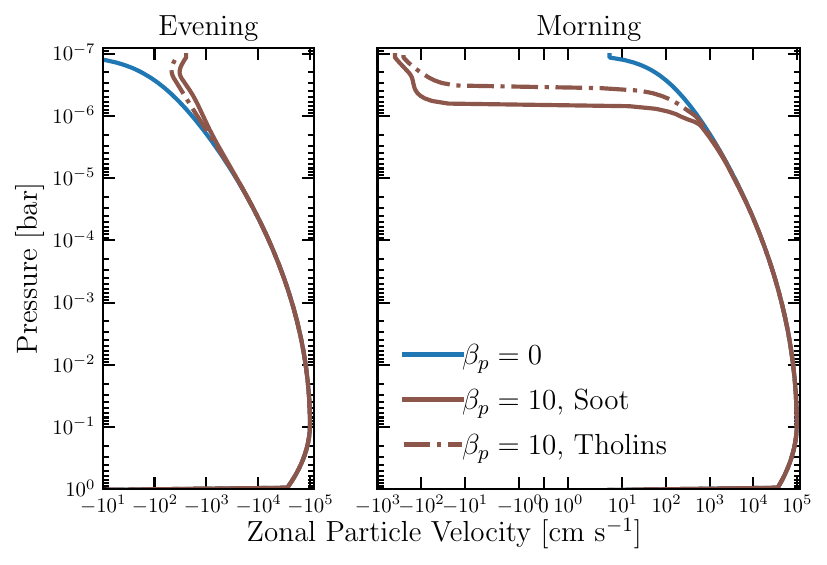}
    \caption{The zonal particle velocities as a function of pressure on the evening (left) and morning (right) terminators for simulations with and without radiation pressure. A negative velocity indicates the flow is towards the night-side. Strong radiation pressure reverses the particle flow at high altitudes on the morning terminator, causing them to flow towards the night-side instead of being advected onto the day-side by the zonal winds.}
    \label{fig:particle_vel}
\end{figure}
The velocity structure is predominately dominated by the zonal wind; however, at high altitudes, it is dominated either by radiation pressure or gravitational settling. We show the zonal particle velocity at the morning and evening terminators in Figure~\ref{fig:particle_vel}. At high radiation pressures, the direction of particle flow at the morning terminator is reversed at high altitudes (also evident in the bottom right panel of Figure~\ref{fig:compare_term}), whereas on the evening terminator, the zonal particle velocity in the upper layers is increased by over an order of magnitude. This reversal/enhancement of the zonal particle velocity in the terminators induces a strong {\rc difference in the optical slopes} of the transmission spectra (Section~\ref{sec:spectra}) between the evening and morning terminators. In fact, high radiation pressures actually result in an outflow of haze particles. For the $\beta_p=10$ simulation, we measure mass-loss rates of  $8.5\times10^{7}$ and $6.6\times10^5$~g~s$^{-1}$ for the soot-like and tholin-like hazes, respectively. The significantly higher mass-loss rates for the soot-like hazes arise from higher opacities for small particles (due to the lower Rayleigh index), which cause larger accelerations on the smaller particles. In particular, the mass-loss rates of the hazes are potentially high enough to be evolutionarily important (see Section~\ref{sec:discuss}).

\section{Transmission Spectra}\label{sec:spectra}
With our 2D equatorial band models, we can now consider the impact of radiation pressure on transmission spectra. Since the signal-to-noise of a transmission spectrum increases with higher irradiation levels and weaker gravities (larger scale heights), as radiation pressure is also more important for planets with higher irradiation levels and weaker gravities, then better targets for transmission spectroscopy will be more impacted by radiation pressure\footnote{Although $\beta_p$ does not scale exactly with the transmission spectroscopy metric (TSM) \citep{Kempton2018}, planets with larger TSM are more likely to have higher $\beta_p$.}. 

\subsection{Method}
For our initial exploration here, we ignore the feedback that the haze particles have on the thermal and chemical properties of the underlying atmosphere (consistent with the assumptions in our simulations). To construct our transmission spectra, we use {\sc petitRadTrans} \citep{Molliere2019} to compute a chemical equilibrium aerosol-free spectrum for our HAT-P-65 b-like analogue with a constant temperature profile, set to the equilibrium temperature. We use the H$_2$O line list from \citet{Water}, the CO line list from \citet{HITEMP}, the CH$_4$ line list from \citet{Methane}, the CO$_2$ line list from \citet{CO2} and collision-induced opacities from \citet{CIA}. We have modified {\sc petitRadTrans} to output the wavelength-dependent optical depths as a function of {\rc the distance from the centre of the planet to the transmission optical path at $Z=0$} ($b$) for the aerosol-free atmosphere ($\tau_{\lambda,{\rm \,gas}}(b)$). 

Using our ray construction from the simulations (Section~\ref{sec:numerics}), we then ray-trace our simulation from the star at different wavelengths, only considering the impact of the haze particles. The procedure then provides the wavelength-dependent optical depths as a function of impact parameter for hazes only ($\tau_{\lambda,{\rm \,haze}}(b)$).

The total transmission optical depth as a function of wavelength and impact parameter is then:
\begin{equation}
    \tau_{\lambda \,{\rm, tran}}(b) = \tau_{\lambda,{\rm \,gas}}(b) +  \tau_{\lambda,{\rm \,haze}}(b)
\end{equation}
Given that our simulation is in 2D, we must make some assumptions about how to distribute the particles in 3D. We follow \citet{Kempton2017} and assume that the aerosol distribution is uniform on the morning and evening terminators, respectively. Specifically, we take our 2D evening terminator distribution and assume it is angularly symmetric for all latitudes on the evening terminator. For the morning terminator, we take our 2D-morning terminator distribution and distribute it in the same manner on the planet's morning terminator. Given that JWST has provided the opportunity to measure separate transmission spectra from the morning and evening terminators \citep{Espinoza2024}, this procedure allows us to perform a similar analysis and calculate the total transmission spectra. 

\subsection{Results}

\begin{figure*}
\centering
\includegraphics[width=\textwidth]{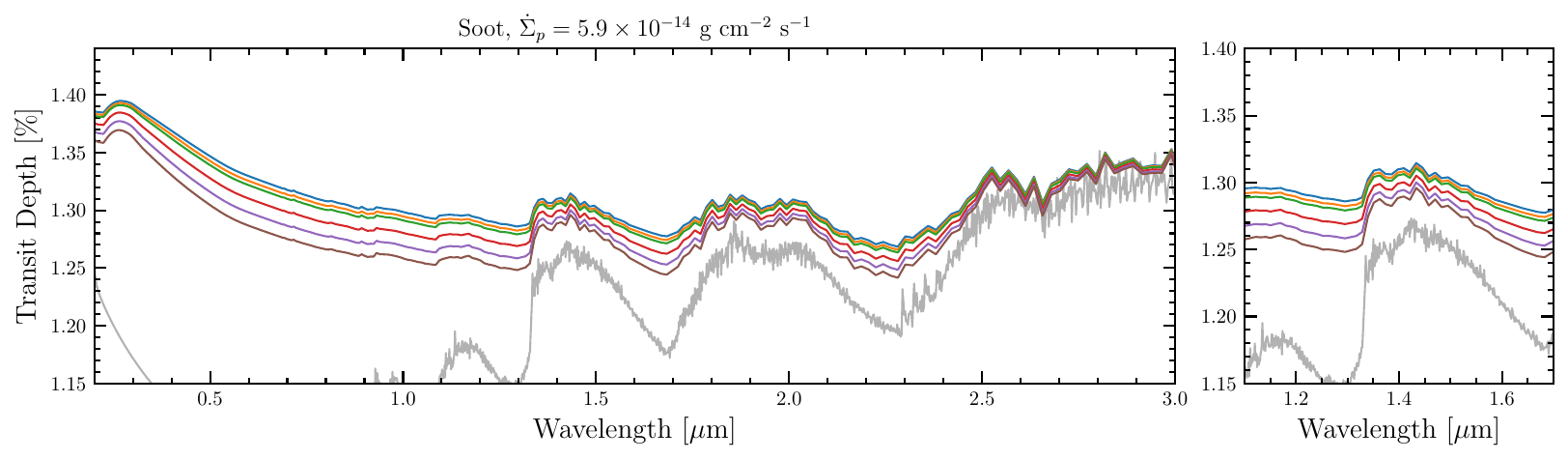}
\includegraphics[width=\textwidth]{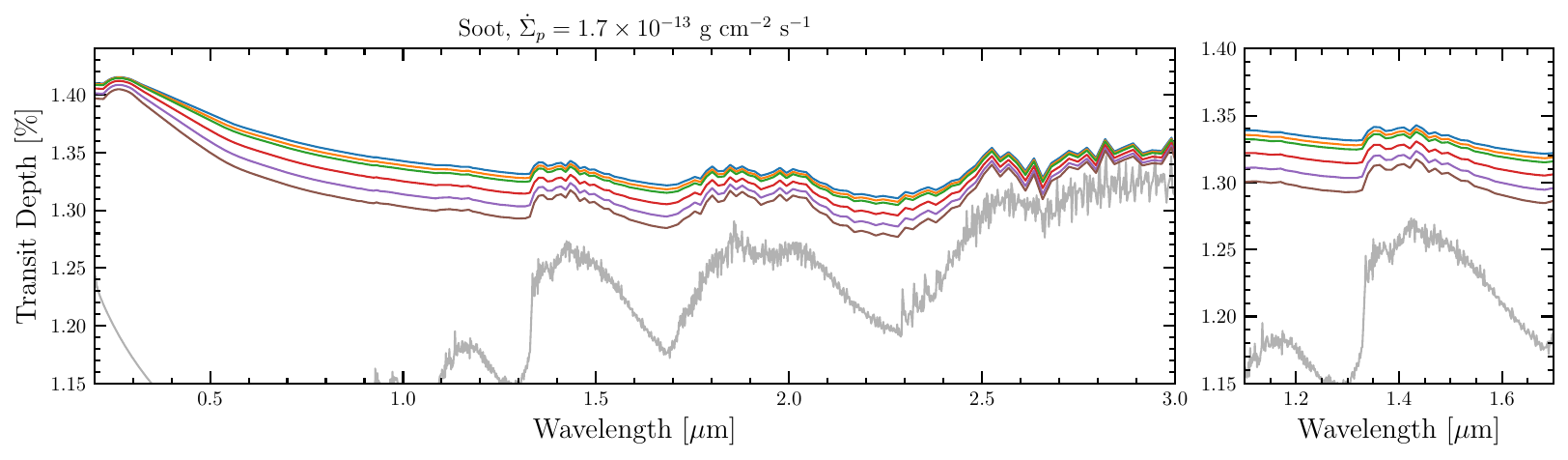}
\includegraphics[width=\textwidth]{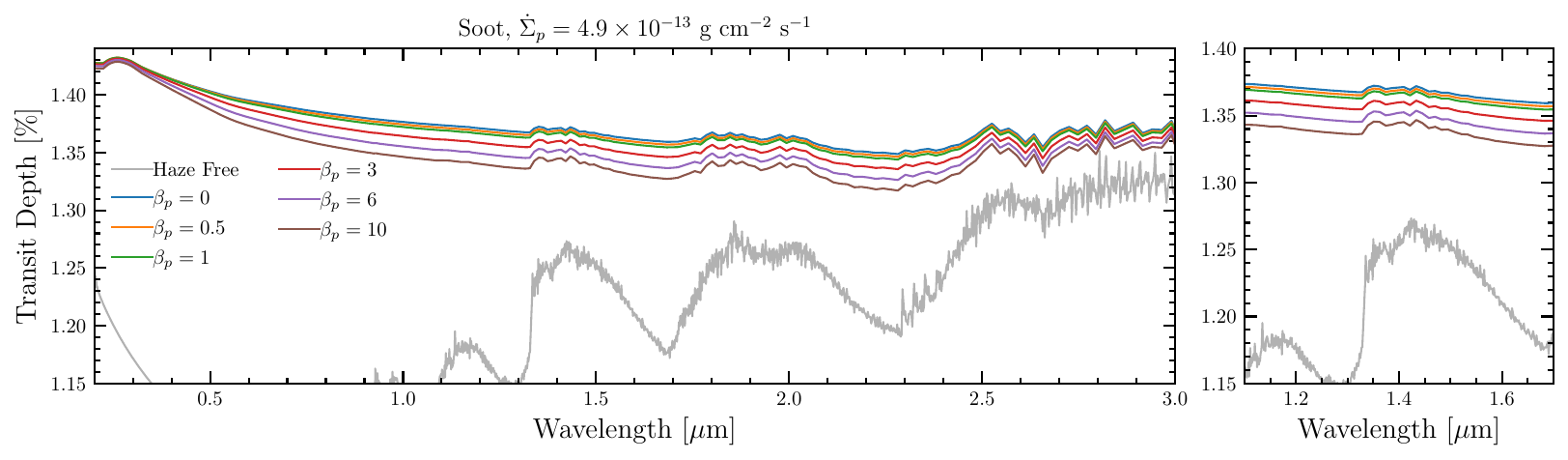}
\caption{The optical-NIR transmission spectrum for our HAT-P-65b analogue with soot-like hazes. The left panel shows the full wavelength range, while the right panel shows a zoom-in on the 1.4$\mu$m water feature. Radiation pressure produces atmospheres that appear less hazy and produce steeper slopes in the optical. The panels show an increasing haze production rate from top to bottom.}
\label{fig:tran_spec}
\end{figure*}

\begin{figure*}
\centering
\includegraphics[width=\textwidth]{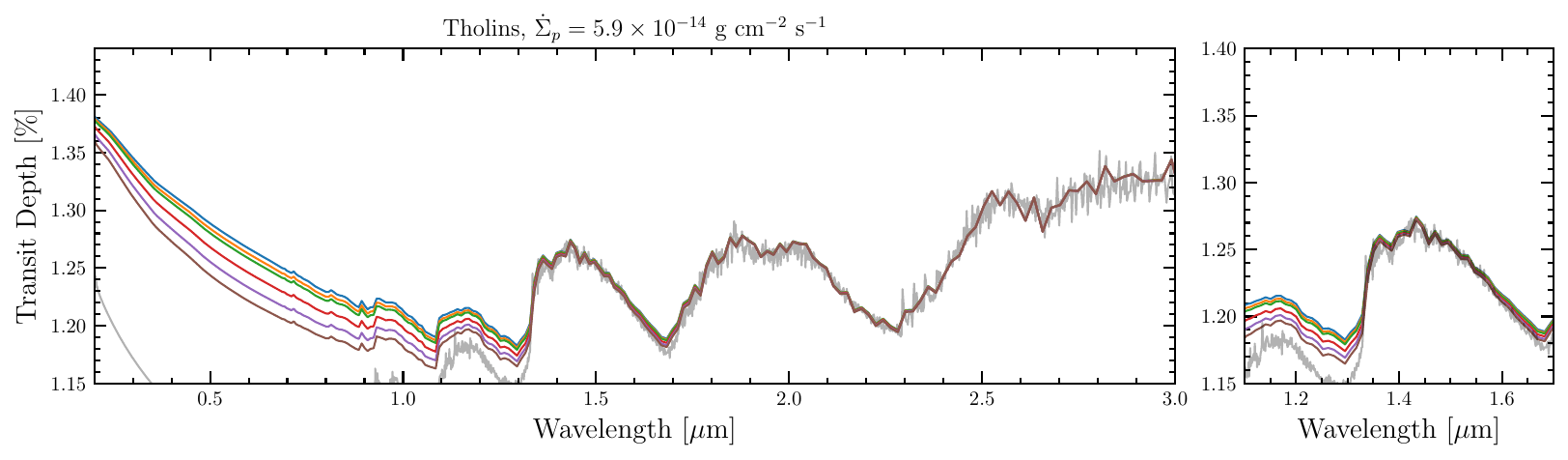}
\includegraphics[width=\textwidth]{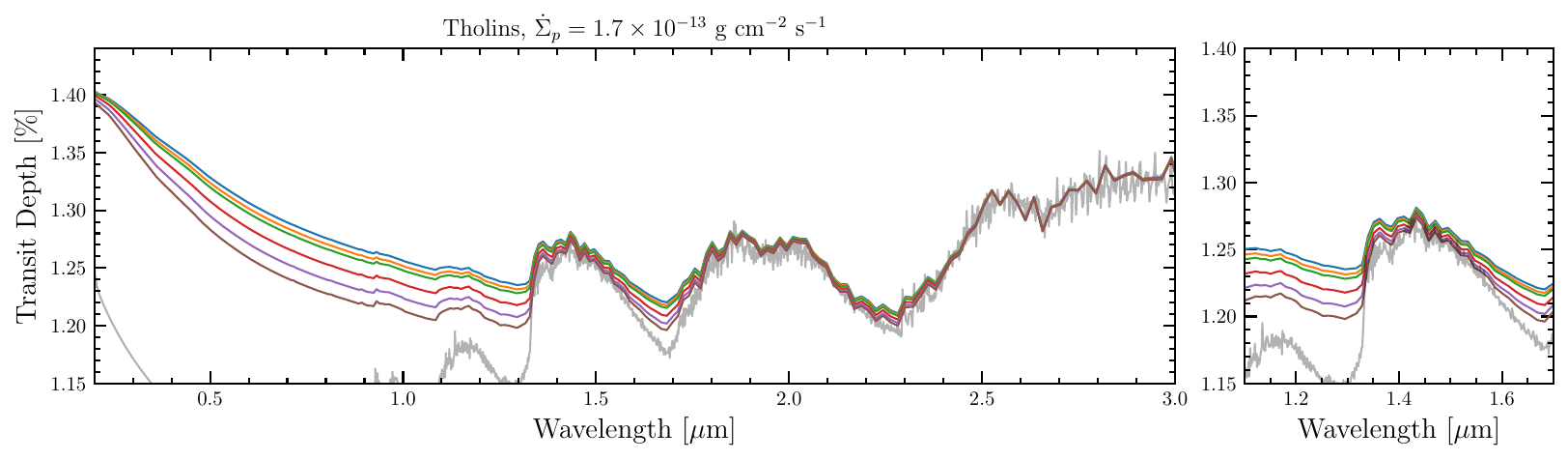}
\includegraphics[width=\textwidth]{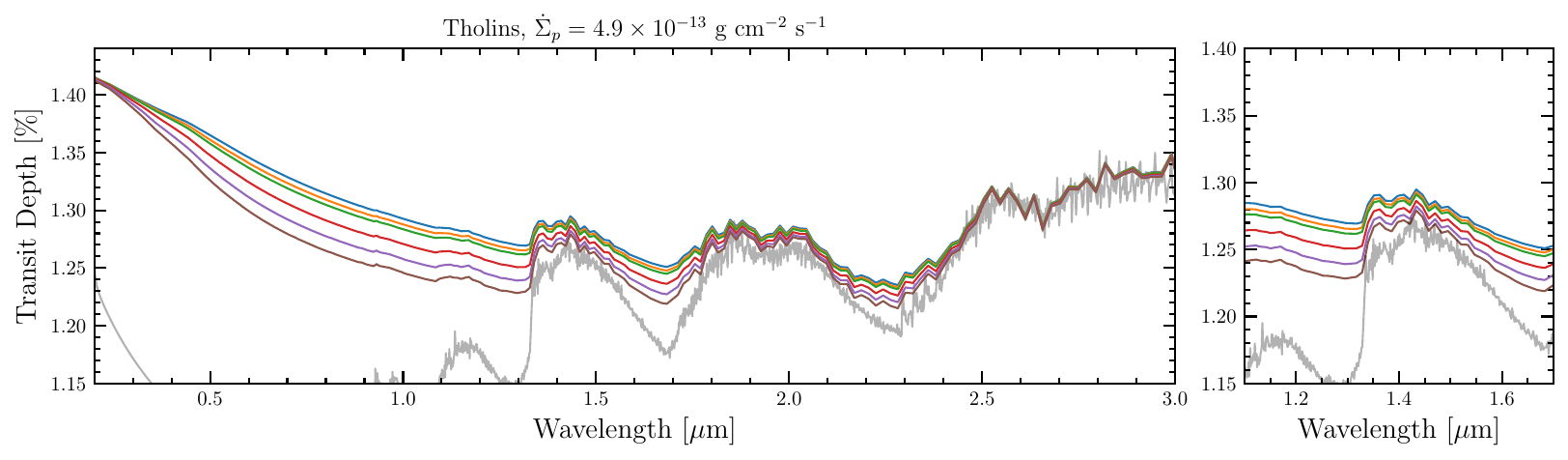}
\caption{Same as Figure~\ref{fig:tran_spec}, but for tholin-like hazes.}
\label{fig:tran_spec_th}
\end{figure*}

In Figures~\ref{fig:tran_spec} and \ref{fig:tran_spec_th}, we show the optical to NIR transmission spectrum of our HAT-P-65 b analogue for different strengths of radiation pressure and haze production rates for both soot- and tholin-like hazes. We adopt HAT-P-65 b's stellar parameters from \citet{Hartman2016HAT} to compute the transmission spectra. These figures also show a zoom-in around the 1.4$\mu$m water feature. As expected from previous work on the impact of aerosols on transmission spectra \citep[e.g.][]{Wakeford2015}, haze particles introduce a scattering slope in the optical wavelength range, out to $\sim$ 1$\mu$m, whereas in the NIR, the aerosols ``mute'' the molecular absorption features (\citealt{Fortney2005}, in this case predominately water). Furthermore, high haze production rates result in larger particles at a given altitude and larger impact parameters where the transit optical depth at a given wavelength is unity, producing shallower scattering slopes in the optical and more muted molecular features in the NIR. Soot-like hazes produce flatter, more muted spectra at a fixed production rate simply due to their lower Rayleigh index, meaning smaller particles have a higher opacity (Figure~\ref{fig:efficiency}). The overall higher opacity mutes the spectrum, while the reduced variation in opacity with wavelength flattens the slope.

However, radiation pressure also plays an important role in shaping the spectra. These figures demonstrate our inference from looking at the particle size and density distribution: that the smaller particles and lower densities (at fixed production rate) from higher radiation pressure result in transmission spectra that appear less hazy and have steeper optical scattering slopes. The impact on the spectral slope is more pronounced at higher production rates as the particles can grow larger due to the higher densities, where they will be more strongly impacted by radiation pressure, particularly in the case of the tholin-like hazes. Therefore, radiation pressure can induce steeper spectral slopes than would be purely expected from the Rayleigh index of the particles alone (i.e. \citealt{Wakeford2015}). Given that our production rate is independent of the strength of the diffusive mixing ($K$, see Section~\ref{sec:discuss}), our results are independent of its strength. Only when the value of $K$ is extremely large $K\gg10^{10}$~cm$^2$~s$^{-1}$ do the results fundamentally change. At these large values, diffusive vertical transport dominates over settling, resulting in haze {\rc mass} concentrations that rapidly decrease with altitude, and we essentially recover the results of \citet{Ohno2020}.

\begin{figure*}
    \centering
    \includegraphics[width=0.495\textwidth]{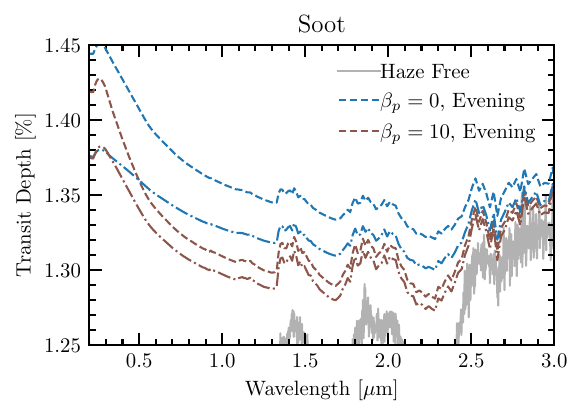}
    \includegraphics[width=0.495\textwidth]{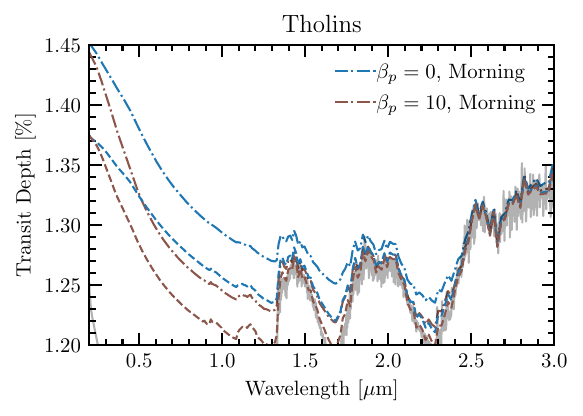}
    \caption{The transmission spectra are shown separately for the evening (dashed) and morning (dot-dashed) terminators, with strong radiation pressure ($\beta_p=10$) and without radiation pressure. The left plot shows soot-like hazes, while the right plot shows tholin-like hazes. Haze-free spectra are shown for scale. Radiation pressure induces steeper optical scattering slopes and less hazy spectra at NIR wavelengths. {\rcc The haze production rate is $1.7\times10^{-13}$~g~cm$^{-2}$~s$^{-1}$ for soot,  $4.9\times10^{-13}$~g~cm$^{-2}$~s$^{-1}$ for tholins, and $K=10^8$~cm$^2$~s$^{-1}$. At these respective production rates the changes in the individual optical slopes due to radiation pressure are particularly large.}}
    \label{fig:limbs}
\end{figure*}

Since radiation pressure has produced particle velocity asymmetry in the upper regions of the atmosphere, we expect there to be an impact on the transmission spectra measured in each limb separately. As shown in Figure~\ref{fig:limbs}, this is indeed what we find, where the evening terminator is hazier (due to the transport of material from the day-side) as expected \citep{Kempton2017}. Furthermore, we also find steeper optical slopes on each terminator separately than without radiation pressure, with the steeper slope more pronounced on the morning terminator (becoming super-Rayleigh, whereas the evening remains sub-Rayleigh); this arises from the velocity reversal of the hazy particles at high altitudes on the morning terminator.  As haze particles are transported toward the night side on the morning terminator by radiation pressure, they are moved to regions with lower haze density, growing slower, resulting in a slower increase in particle size and density with depth and hence opacity that falls faster with increasing pressure, producing a steeper optical slope. However, we find the absolute difference in the transit depths between the two terminators is reduced by radiation pressure.  In Section~\ref{sec:discuss}, we speculate that this could be an important process for cloudy atmospheres that are expected to have cloudier morning rather than evening terminators \citep[e.g.][]{Kempton2017}. 

\section{Discussion}\label{sec:discuss}

\begin{figure}
    \centering
    \includegraphics[width=\columnwidth]{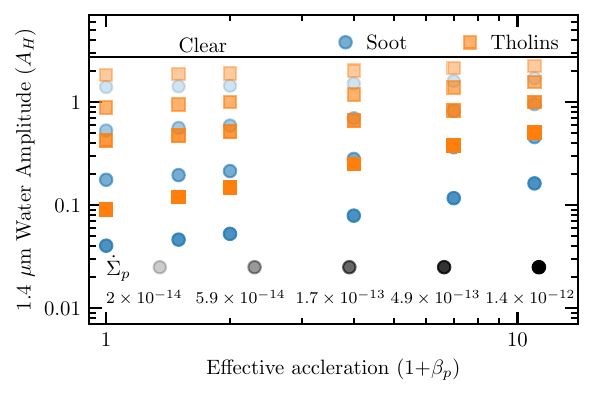}
    \caption{The normalised amplitude of the 1.4 $\mu$m water feature (Equation~\ref{eqn:water_amp}) as a function of the total acceleration experienced by particles ($1+\beta_p$) and haze production rate in g~cm$^{-2}$~s$^{-1}$ for our HAT-P-65b analogue. The black line represents the value of a completely clear atmosphere. At higher accelerations, the haze particles move faster and have lower densities (at fixed production rate), resulting in clearer transmission spectra and, thus, larger normalised amplitudes of spectra features. Thus, if radiation pressure is a dominant controlling factor for aerosols on hot exoplanets, spectral feature amplitudes should increase with the strength of radiation pressure. The lowest and highest haze production rates are not shown for the tholins (essentially clear) and soot (essentially haze-dominated), respectively.  }
    \label{fig:water_amplitude}
\end{figure}
We have demonstrated that stellar radiation pressure on aerosol particles is dynamically important for close-in, highly irradiated planets. For aerosols with sizes $\sim 0.1~\mu$m, the magnitude of the acceleration due to radiation pressure can exceed ten times the planet's gravity for the most highly irradiated exoplanets. Radiation pressure not only impacts the dynamics of the aerosols in the atmosphere but also changes its observable properties in terms of the planet's albedo and its transmission spectrum. In particular, all other parameters being equal, radiation pressure produces faster-moving particles producing lower aerosol {\rc mass} concentrations, resulting in smaller particle sizes and, therefore, lower albedos with steeper optical slopes in transmission and {\rc larger amplitude NIR molecular features}. 

\subsection{Model Limitations}
In this work, we have considered the impact of stellar radiation pressure on aerosol particles in close-in exoplanet atmospheres for the first time. Thus, to isolate the physics, we have made numerous assumptions and simplifications worth discussing. In this work, we have exclusively focused on ``haze'' particles (i.e., those that do not grow by condensation) and simply coagulate after production.  Firstly, we have parameterised the haze production rate, both in absolute strength, functional form, and initial size. These choices were motivated by previous works based on more complete models \citep[e.g.][]{Gao2018,Kawashima2019}; however, it is worth discussing the implications of these choices. In particular, since hazes are believed to be photochemically produced \citep{Gao2021}, which correlates with the strength of UV flux that the planet experiences \citep[e.g.][]{Yu2021}, there will clearly be a (implicit) correlation between haze production rate and the strength of the radiation pressure, which can have observable impacts on Albedo trends (Figure~\ref{fig:albedo_obs}). Furthermore, the haze precursor species (e.g. methane) also correlate with the planet's equilibrium temperature, introducing further correlations between the haze production rate and the strength of radiation pressure. Thus, further work including radiation pressure in more complete haze production models \citep[e.g.][]{Helling2008,Gao2018} must be considered before the exact dependence of any correlation with radiation pressure can be compared to observations.

We have assumed plane parallel ray tracing from the star to calculate the radiation pressure force on the aerosol particles. While plane parallel rays are clearly a good assumption given the ratio of the planet's size to its distance from the host star (even for the most highly irradiated ones), we have neglected two other radiation fields. All the incident stellar irradiation is radiated back out of the atmosphere by the planet (assuming global thermodynamic equilibrium), and the planet additionally radiates out a small internal luminosity. Given {\rc that the intrinsic temperatures} of typical planets (even when young or inflated) are at maximum, a few 100~K, this internal luminosity is clearly unimportant. However, the re-radiated stellar light must be equal in energy budget to the input. The re-radiated stellar light is either thermally emitted or scattered, both of which apply a net radiation pressure force in the opposite direction to the planet's gravity. However, while the stellar radiation field is plane-parallel, the thermally emitted and scattered radiation field is more diffusive, as it is emitted over a larger range of solid angles.

Assuming a plane-parallel atmosphere above the photosphere, the ratio of the radiation pressure from these diffusive fields to a plane-parallel one is $1/2$ for thermal emission, whereas it is $1/\sqrt{3}$ for the scattered radiation field, assuming isotropic scattering \citep{Chandrasekhar1960}. This compares to $1$ for the incoming radiation field (at the sub-stellar point). Thus, the radiation pressure force ratio for the scattered stellar light, compared to the incoming stellar light, is $\sim R/\sqrt{3}$. Using the evaluation of Equation \ref{eqn:R} displayed in Figure~\ref{fig:albedo}, when $\beta_p>1$, this ratio is at maximum $\sim 0.15$ for high albedos. Thus, while not entirely negligible, the scattered radiation field is certainly not important. However, the thermal field is negligible: for small particles, the ratio of the radiation pressure force from the re-radiated thermal radiation to the stellar radiation field is $\sim (1-R)(T_{\rm eq}/T_*)^{(1+b)}/2$ assuming no atmospheric heat redistribution, which is at maximum only a few percent. Atmospheric heat redistribution by winds will lower this even further. Thus, future work might consider the possible $\sim$$10\%$ correction to the radiation pressure force that arises from scattered light by the aerosols themselves. 

Furthermore, we have ignored the impact of radiation pressure on the gas. Collisions between aerosol particles and gas will transmit the deposition of radiative momentum from the particles to the gas as they accelerate to the terminal velocity. However, the gas feels an acceleration a factor of $X_p$ times the radiative acceleration felt by the aerosols. Since $\beta_p$ is a maximum of a few 10s, the {\rc mass} concentration of aerosols would need to be higher than a few percent for the radiation pressure to have any impact $\sim 10\%$ on the gas. Concentrations of this high level require extremely high production rates; thus, we suspect that radiation pressure on the gas is rarely important. 

As our starting point, we have adopted compact spherical grains in our opacity model. However, coagulation does not have to result in compact grains of larger sizes; coagulation can result in ``fluffy'' aggregates with a high fractal dimension. Indeed, such fractal growth has been shown to change the observed properties of aerosols in an exoplanet's atmosphere \citep[e.g.][]{Adams2019,Vahidinia2024}. Furthermore, we expect that fractal growth will lead to a strong change in the response of the particles to the impact of radiation pressure. This is because the scattering opacity grows non-linearly with increasing monomer number in the aggregate, whereas the absorption opacity grows only linearly \citep[e.g.][]{Sorensen2001}. Thus, future work should explore the role of radiation pressure on fractal aerosols. 

In terms of our underlying atmosphere model, in this work we simply assumed a global isothermal atmosphere that is invariant to the properties of the aerosols in the atmosphere. It is well established that the atmosphere is not globally isothermal, both with altitude and position on the planet, and there is now good observational evidence that the morning and evening terminators have different temperatures \citep{Espinoza2024}. Our isothermal assumption, therefore, means that we cannot study the thermal feedback between radiation pressure and the temperature structure of the atmosphere. For example, thermal feedbacks may exist: higher radiation pressure results in smaller particle sizes and lower {\rc mass} concentrations, which lowers the atmosphere's opacity to stellar irradiation. Lower opacities (and lower albedos) result in more stellar irradiation being absorbed deeper into the atmosphere and higher atmospheric temperatures. Higher atmospheric temperatures increase the scale height, increasing the settling time and hence promoting growth and larger particles. This might have two opposite effects depending on the particle's Rayleigh index. For particles with Rayleigh indices close to zero (e.g., soots), the radiation pressure opacity is roughly independent of the particle size for small particles (Figure~\ref{fig:efficiency}). As such, the higher temperatures would increase the particle concentration (Equation~\ref{eqn:conc_arad}), creating a negative feedback. Alternatively, for particles with a non-zero Rayleigh indices (e.g. tholin-like hazes and silicates), the slower settling time would increase the particle size, resulting in higher radiation pressures, and could produce lower aerosol densities, increasing the temperature further. Thus, in the case of particles with large Rayleigh indices, a positive feedback could result, producing time-varying (or bi-stable) aerosol distributions. Thus, this feedback should be further investigated. 

One obvious drawback of our isothermal assumption is that we are unable to calculate emission spectra; however, we speculate, based on the above discussion, that radiation pressure would result in hotter atmospheres.  Furthermore, since the emission spectrum can provide some constraints on the temperature gradient in the atmosphere \citep{Kreidberg2018}, as we have modified the opacity structure giving rise to a pressure dependence that is different in the case of radiation pressure (Equation~\ref{eqn:opac1}) compared to without, then the temperature gradient would be modified which would potentially modify the emission spectra. Thus, in the future, stellar radiation pressure should be considered in models that actively consider the atmosphere's thermodynamics. 

Given that stellar radiation pressure does not act in a spherically symmetric manner on a tidally locked exoplanet (Figure~\ref{fig:2d_schematic}) we have also chosen to model the impact of radiation pressure on a 2D zonal equatorial band. Since atmospheres are 3D, we have assumed that the 2D zonal equatorial band is symmetric on its respective morning or evening terminator to compute transmission spectra. Furthermore, we have parameterised the equatorial zonal wind with a simple fitting function based on GCM simulation results; however, given the computational expense, we have chosen not to explore varying this functional form. While moving to 3D and using exact wind profiles extracted from GCMs for the exact planet of interest would improve the accuracy of our model, they would not change the basic picture: that zonal winds transport hazes created on the day-side to the evening terminator and that radiation pressure impacts the particle size and concentration distribution. We do note that at extreme irradiation levels, the zonal jets transition to day-to-nightside flows \citep{Tan2019}. This would certainly change our asymmetry results, but the basic impact of radiation pressure would still hold. 

Although we have made many assumptions and simplifications in this initial exploration, the basic physics remains robust. Thus, while we must be cautious in interpreting the exact results from our simulations, the basic result that radiation pressure will produce smaller particles with lower {\rc mass} concentrations should be unaffected by our simplistic approach.  Therefore, our general observational conclusions should be robust and, in particular, qualitative trends should be robust. 

\subsection{Observational implications}
We have demonstrated that radiation pressure has an observable impact on both a planet's albedo and transmission spectra and speculated that it will impact the emission spectrum. The main result is that aerosols move faster as a result of a strong additional acceleration from the radiation pressure. As such, all other factors being equal, the faster-moving aerosols grow slower and have lower {\rc mass} concentrations, resulting in lower aerosol opacity. The lower aerosol opacity generally makes atmospheres appear less ``hazy'' than without radiation pressure. Although microphysical aerosol models without radiation pressure have had some success in explaining general features, such as the amplitude of the 1.4~$\mu$m water feature \citep{Gao2020} they do not explain the full properties of observations, including the fact that some hot Jupiter's day-sides appear clearer than the models predict \citep{Gao2021} thus a possible solution to this is the introduction of radiation pressure. 

Furthermore, there has long been known to be a large diversity in the optical slopes measured in the transmission spectra \citep{Sing2016}, which the microphysical models are yet to explain fully. In particular, very steep slopes have been observed in some exoplanets \citep[e.g.][]{Pinhas2019,Welbanks2019}. \citet{Ohno2020} proposed that very strong vertical mixing due to turbulence could explain these slopes; however, we also find that radiation pressure induces significantly steeper optical slopes. {\rm Thus, given these possible degenerate solutions more work is required to determine the best way to observationally distinguish the importance of radiation pressure from other processes.} This impact could be particularly important in the case of cloud-like particles which might only exist on the morning terminator \citep{Kempton2017};  {\rcc as we show in Figure~\ref{fig:limbs}, the transmission spectra on individual morning and evening terminators are impacted by radiation pressure, with changes to the relative absorption amplitudes and optical slopes. The typical impact is that radiation pressure steepens the optical slope, and the depths of the terminators are different due to zonal transport. However, these changes are sensitive to the type and rate of haze production and need to be studied in more detail, since differences between the morning and evening terminators are now observed \citep[e.g.][]{Espinoza2024,Murphy2025}. }

Currently, reflected light properties (besides the albedo) are difficult to measure for most exoplanets. However, with the next generation of ground-based optical telescopes, such information will become accessible \citep{Gao2021}. Given radiation pressure has altered both the size and concentration distribution, we expect that both the reflected light spectra \citep{Barstow2014} and phase curves \citep{Heng2021} will be impacted. Therefore, future work should consider the impact of radiation pressure on the properties of reflected light from hot exoplanets.

\subsection{Aerosol Mass-loss}

One of the most intriguing results from our simulations is the identification of the loss of aerosol particles from the atmosphere, with radiation pressure driving an outflow of haze particles at certain points in the atmosphere. For soot hazes, we found mass-loss rates approaching $\sim 10^8$~g~s$^{-1}$. Given a typical giant planet with a solar metallicity atmosphere, this mass-loss rate is inconsequential for the evolution of the planet's bulk.  

However, comparisons with the bulk reservoir of the planet may not be appropriate.  For both haze-like and cloud-like aerosols, as they settle deeper into the atmosphere, the new precursor material is mixed up from deeper. In the standard case, there is a detailed balance in which the rate of material settling matches the amount of precursor material mixed up, and as such the abundance of the precursor material deeper in the atmosphere remains constant. However, as we now have aerosol mass-loss, the reservoir of precursor material supplying the region producing the aerosols must be depleted. For our soot-like haze simulation, the mass-loss rate is actually $\sim 1.6$\% of the production rate, meaning that this region can be depleted on $\sim 60$ mixing timescales. Assuming that the aerosol progenitors come from the radiative region and from a depth that makes up a maximum $\sim10\%$ of the planet's total radius, for a $K_{\rm zz}=10^{8}$~cm$^2$~s$^{-1}$, there are $\sim 10^6$ mixing timescales in 5~Gyr. Even for the case of tholin-like hazes our mass-loss rates imply the supply region would be depleted on $\sim 8000$ mixing timescales. Therefore, given that the depletion timescale is considerably shorter than the planet's lifetime, the upper atmosphere could become depleted in haze progenitors over the planet's lifetime.  Alternatively, the radiative atmosphere could be replenished from the convective interior. Therefore, these aerosol mass-loss models should be coupled with atmospheric structure and interior evolution/mixing models to consider their long-term chemical evolution. Such an evolution would be particularly interesting in the context of a sub-Neptune planet which can have radii $\sim 10~$R$_\oplus$ at young ages, implying a $\beta_p$ value in excess of 10 for a typical sub-Neptune with a mass of 5~M$_\oplus$ and separation of 0.1~AU \citep{Owen2020}. Thus, the impact of radiation pressure is an additional confounding variable when trying to interpret the chemical abundances of planets in terms of their formation \citep{Kirk2024,Penzlin2024}.

\subsection{Population level trends}

Understanding population-level inferences of aerosols in exoplanets has remained challenging \citep{Gao2021}, mainly due to limited sample sizes and an incomplete theoretical understanding of aerosol dynamics and their impact on exoplanet spectra. Perhaps one of the most interesting parameters is the amplitude of the 1.4~$\mu$m water feature that has been measured for a sample of $\sim 30$ exoplanets \citep{Fu2017,Dymont2022}; {\rc however, it is important to be cautious interpreting these population studies as they include a significant fraction of cooler sub-Neptunes where radiation pressure will not be important}. \citet{Dymont2022} showed that there was no clear understanding of the trends shown in the data with planetary or stellar parameters, although there were tentative trends with planetary equilibrium temperature and gravities. Higher temperatures and weaker gravities show larger normalised water feature amplitudes. Given that the normalised water feature amplitude scales its strength in terms of the atmospheric scale height, this trend does not simply arise from larger scale heights. Since both equilibrium temperature and planetary gravities have implicit correlations with the relative strength of radiation pressure relative to planetary gravity, we expect the amplitude of gas phase spectral features to correlate with the strength of radiation pressure. In figure~\ref{fig:water_amplitude}, we show that the normalised strength of the 1.4~$\mu$m feature, defined as (and computed identically to \citet{Dymont2022}:
\begin{equation}
    A_H = \frac{R_p(1.4~\mu{\rm m})-R_p(1.25~\mu{\rm  m})}{H}\label{eqn:water_amp}
\end{equation}
We do indeed find a correlation where high radiation pressure produces less hazy atmospheres and strong water absorption amplitudes. This correlation is present for both soot-like and tholin-like hazes and is apparent across a wide range of production rates (essentially, all production rates provide a detectable water feature in a hazy atmosphere). Therefore, we speculate that radiation pressure strength is an additional parameter that controls the as yet unexplained 1.4~$\mu$m  water-feature amplitude's variation with stellar and planetary parameters. We caution that because in our models we have kept the planetary and stellar parameters fixed (including equilibrium temperature), while varying the strength of the radiation pressure, the real correlation between a spectral feature amplitude and the strength of radiation pressure is likely to be different. This is because, as discussed earlier, the strength of radiation pressure implicitly correlates with both planetary equilibrium temperature and stellar UV flux (and hence the production rate, \citealt{Horst2018,Kawashima2019}), both of which impact the haze properties (Equations~\ref{eqn:conc_arad} and \ref{eqn:size_arad}). Therefore, while we expect a positive correlation between the amplitude of a spectral feature and the strength of radiation pressure, we cannot predict the details of this correlation without a considerably larger simulation set. Indeed, if the haze production rate was to correlate linearly with (bolometric) flux, then we could significantly flatten the trend; alternatively, if the haze production rate were to decrease with increasing equilibrium temperature (due to reduction in haze precursor molecules, e.g. \citealt{Kawashima2019}) then this would steepen any trend. Furthermore, this assumes that the aerosols obscuring the water feature are photochemically produced hazes; rather, if they are cloud condensates, then complex dependencies on atmospheric structure become important.

\subsection{Condensate clouds}
While we have focused initially on photo-chemically produced hazes due to their slightly simpler modelling, condensate clouds are thought to be present in many hot exoplanets with silicates the likely candidate \citep[e.g.][]{Gao2020}. Indeed, JWST  has now observed silicate clouds \citep{Grant2023}. One of the reasons we've chosen to study hazes over such a broad region of parameter space (including regions where microphysical models suggest they might not be present) is because the basic physics of aerosol dynamics translates from haze particles to cloud particles.  We expect similar impacts on their dynamics -- faster-moving particles -- resulting in smaller particles and lower {\rc mass} concentrations. Thus, we would expect a similar effect on the transmission spectrum, namely steep optical slopes. {\rc Due to condensation, cloud particles may be larger than the hazes we have modelled for the same production rate. If these particles are larger than the typical wavelength of star-light, where $Q_{\rm rad}$ is approximately independent of particle size, their dynamics and observational implications maybe different from those found here.} Furthermore, since clouds might be more strongly confined to the morning terminator than our haze models, which impact both terminators (albeit not equally), we suspect the very steep (super-Rayleigh) scattering slopes induced by radiation pressure on the morning terminators in our haze model Figure~\ref{fig:compare_term} could be the explanation for some of the super-Rayleigh scattering slopes observed in hot Jupiters \citep[e.g.][]{Pinhas2019,Welbanks2019}, when combined with an aerosol-free evening terminator (that would yield a Rayleigh scattering optical slope). However, future modelling of cloud particles needs to be performed to confirm this speculation.  


\section{Conclusion}

The impact of aerosol particles on the observations of exoplanet atmospheres has been clear for over a decade \citep[e.g.][]{Deming2013,Sing2013}. However, the physical processes that govern their properties remain poorly understood \citep[e.g.][]{Helling2019,Gao2021}. In this work, we show that radiation pressure from the host star's irradiation can be dynamically important, often for the most highly irradiated planets dominating over the acceleration due to the planet's gravity. With ratios of the acceleration arising from radiation pressure compared to the planet's gravity exceeding $\sim 10$ for common aerosols with particle sizes $\sim 0.1~\mu$m. 

We have focused on the impact of radiation pressure on haze-like particles and simulated the impact on their dynamics in 2D, tracking a tidally locked planet's equatorial band. Using our models and simulations, we have also computed the impact of increasing radiation pressure on observational properties of the atmosphere. Our main findings are as follows:
\begin{enumerate}
    \item Aerosols with a size of 0.1 $\mu$m have ratios of radiation pressure to gravitational acceleration that exceed unity for $\sim 52$ (soot-like hazes) and $\sim 24$~\% (Mg$_2$SiO$_4$ condensates) of known gas-rich exoplanets with measured masses and radii. Furthermore, since radiation pressure is more important for planets which are better targets for atmospheric characterisation, radiation pressure is likely to be an important process on aerosols for many planets with observable atmospheres. 
    \item Radiation pressure increases the acceleration felt by aerosols, causing them to move through the atmosphere at higher speeds. This results in slower growth and lower {\rc mass} concentrations. 
    \item For hazy atmospheres, radiation pressure significantly reduces the production rate range over which the particles are numerous but small enough at high altitudes to give rise to a high albedo. 
    \item Radiation pressure and zonal winds have a {\rc difference} between the morning and evening terminators in the direction in which they move the particles, acting in concert at the evening terminator and in the opposite direction at the morning terminator.  {\rcc These processes impact the transmission spectra of individual morning and evening terminators. }
    \item A high radiation pressure, haze particles can be lost from the atmosphere. The mass-loss rates are sufficiently high that haze precursors could be depleted from the radiative region of the atmosphere unless they are replenished sufficiently quickly from the interior. 
    \item The smaller particles and lower {\rc mass} concentrations result in atmospheres that appear less hazy in transmission, giving rise to larger molecular feature amplitudes in the NIR. In addition, the slope of the optical transmission spectrum becomes steeper in the presence of radiation pressure, and the effect is particularly pronounced in the individual morning and evening terminators. This could potentially be an explanation for the steep optical transmission spectra measured in some hot Jupiters. 
    \item Molecular feature amplitudes and scattering slopes correlate with the strength of radiation pressure; however, because radiation pressure also correlates implicitly with other planetary and stellar parameters (e.g. equilibrium temperature, stellar effective temperature, haze production rate), the quantitative predictions are unclear until the links between all the parameters are understood. 

    \item We speculate that radiative feedback between the impact of radiation pressure on aerosols and the atmosphere's temperature could produce both positive and negative feedback loops that warrant further investigation. 
\end{enumerate}

In summary, the impact of radiation pressure on aerosols could be one of the missing pieces in understanding their physics in exoplanet atmospheres and could be the explanation for much of the outstanding scatter that exists in interpreting the impact of aerosols on the observations \citep[e.g.][]{Gao2021,Dymont2022}.

\section*{Acknowledgements}
We are grateful to the anonymous referee for comments and suggestions that improved the manuscript. The authors thank Beatriz Campos Estrada for comments on an earlier version of the manuscript.
JEO is supported by a Royal Society University Research Fellowship. This project has received funding from the European Research Council (ERC) under the European Union’s Horizon 2020 research and innovation programme (Grant agreement No. 853022). RMC acknowledges support from NASA'S Interdisciplinary Consortia for Astrobiology Research (NNH19ZDA001N-ICAR) under grant number 80NSSC21K0597. This work benefited from the 2022, 2023 and 2024 Exoplanet Summer Program in the Other Worlds Laboratory (OWL) at the University of California, Santa Cruz, a program funded by the Heising-Simons Foundation. This work was performed using the Cambridge Service for Data Driven Discovery (CSD3), part of which is operated by the University of Cambridge Research Computing on behalf of the STFC DiRAC HPC Facility (www.dirac.ac.uk). The DiRAC component of CSD3 was funded by BEIS capital funding via STFC capital grants ST/P002307/1 and ST/R002452/1 and STFC operations grant ST/R00689X/1. DiRAC is part of the National e-Infrastructure. This research has made use of the NASA Exoplanet Archive, which is operated by the California Institute of Technology, under contract with the National Aeronautics and Space Administration under the Exoplanet Exploration Program. For the purpose of open access, the authors have applied a Creative Commons Attribution (CC-BY) licence to any Author Accepted Manuscript version arising.

\section*{Data Availability}
The simulation code is freely available on github: \url{https://github.com/jo276/2Datmosparticle}



\bibliographystyle{mnras}
\bibliography{ref} 







\bsp	
\label{lastpage}
\end{document}